\documentclass[aps,prd,amsfonts,amssymb,amsmath,mathrsfs,nofootinbib,reprint,showpacs,fixltx2e,subfig,usegraphicx,twocolumn]{revtex4}
\usepackage{graphicx}
\usepackage[font=small,labelfont=bf,textfont=footnotesize,justification=raggedright,singlelinecheck=true]{caption}
\usepackage{subfig}
\usepackage{aas_macros}
\usepackage{mathrsfs}
\usepackage{comment}
\usepackage{appendix}
\usepackage{multirow}
\usepackage{url}
\usepackage{color}
\usepackage[normalem]{ulem}
\usepackage{cancel}

\newcommand{\jon}[2]{{#2}}
\newcommand{\joned}[1]{}

\newcommand{\incgraph}[3]{\includegraphics[angle=#1, width=#2\textwidth]{#3}}

\begin{document}

\title{Searching For Anisotropic Gravitational-wave Backgrounds\\ Using Pulsar Timing Arrays}

\author{Stephen R. Taylor}
\email[email: ]{staylor@ast.cam.ac.uk}
\author{Jonathan R. Gair}
\email[email: ]{jgair@ast.cam.ac.uk}
\affiliation{Institute of Astronomy, \jon{}{University of Cambridge}, Madingley Road, Cambridge, CB3 0HA, UK }


\date{\today}

\begin{abstract}
We present the results of simulated injections testing the first Bayesian search-pipeline capable of investigating the angular-structure of a gravitational-wave (GW) background influencing pulsar signals. A stochastic background of GWs from the incoherent superposition of many inspiraling supermassive black hole binaries at nHz frequencies is likely to be the dominant GW signal detectable by pulsar timing arrays (PTAs). Even though one might expect a background composed of a high-redshift cosmological population of sources to be fairly isotropic, deviations from isotropy may be indicative of local GW hotspots or some form of continuous anisotropy in the angular-distribution of GW-power. A GWB induces time-of-arrival deviations in pulsar signals which are correlated between separated pulsars. In an isotropic background this cross-correlation follows a distinctive relationship, known as the Hellings and Downs curve, that depends only on the angular separation of the pulsars. If the background is anisotropic, the cross-correlation is different, but predictable, and also depends on the absolute position of the pulsars. By simulating datasets containing GWBs with various anisotropic configurations, we have explored the prospects for constraining anisotropy using near future data. We find that at moderate to high signal to noise ratio the assumption of isotropy is no longer an appropriate description of the simulated background. Furthermore, we can recover the nature of the injected anisotropy in a Bayesian parameter-estimation search, and propose a prior on the anisotropy search-space motivated by the physicality of the implied distribution of sources. 
\end{abstract}

\pacs{}

\maketitle

\section{Introduction}
The next several years should see the true beginnings of observational gravitational-wave (GW) astronomy. The ground-based interferometers, such as AdLIGO \citep{AdvLIGO}, AdVirgo \citep{AdvVirgo} and KAGRA \citep{kagra2012}, are  expected to be operating at full design sensitivity by the end of this decade, although the first science runs may occur several years before that. These detectors will be sensitive to the chirping signal of stellar mass compact-binary inspirals in the frequency range $\sim10-10^3$ Hz. Additionally, it is hoped that a space-based interferometer with arm-lengths of $\sim10^9$ m, such as eLISA/NGO \citep{elisa-ngo,elisa_science_theme}, will be operable by the end of 2020s, and sensitive to GWs in the frequency range $\sim 0.1-100$ mHz generated by binaries of massive black holes (MBHs) and the extreme-mass-ratio inspirals of stellar mass compact  objects into MBHs.

Gravitational waves can also be detected through their effect on pulsar timing residuals. Several recent studies have claimed that pulsar timing arrays (PTAs) \citep{foster-backer-1990} are on the cusp of detecting a nHz stochastic GW-background \citep{mcwilliams_imminent,mcwilliams_stalled_mergers,sesana_systematic_signal}. PTAs exploit the timing accuracy of millisecond pulsars \citep{pulsar-precision-1,pulsar-precision-2} to measure the influence of a GW perturbing the space-time metric along each pulsar-Earth line-of-sight \citep{sazhin-1978,detweiler-1979,estabrook-1975,burke-1975}. The frequency range that a PTA is sensitive to is set by the pulsar observational time-span ($f_{\rm low}\sim1/T$), and the regularity of pulsar observations ($f_{\rm high}\sim1/(2\Delta T)$). The pulsar array essentially forms a kpc GW-detector which is complementary to other GW experiments.

In practice we must disentangle the influence of a GW on the time-of-arrival (TOA) of pulses from other stochastic and deterministic influences. These include the quadratic spindown of the pulsar itself, as its rotational energy is extracted to power the EM outflow. There are also a variety of noise sources, such as clock noise, receiver noise and dispersion measure effects. The deterministic parameters can be fit out of the raw TOAs to leave the timing-residuals, which include all unmodelled phenomena, such as GWs.

The dominant GW signal expected in the PTA band will result from the incoherent superposition of signals from inspiraling SMBHBs, creating a stochastic background. Essentially many signals emit GWs up in the same frequency bin such that, at low frequency at least, no single source can be individually resolved. At nHz frequencies this creates a characteristic strain-spectrum that is approximately a power-law~\citep{begelman1980,phinney2001,jaffe-backer-2003,wyithe-loeb-2003}. However, at higher frequencies ($10^{-8} - 10^{-7}$ Hz), the signal may become dominated by several close, bright binaries, such that the stochasticity of the signal breaks down \citep{sesana-vecchio-colacino-2008} and the shape of the spectrum will show a stronger dependence on the particular MBH assembly scenario in our Universe.

The characteristic strain-spectrum of a GWB is defined in terms of the fractional contribution of the background to the energy-density of the Universe,
\begin{equation}
\Omega_{\rm GW}(f) = \frac{1}{\rho_{\rm{c}}}\frac{d\rho_{\rm GW}(f)}{d(\ln f)} = \frac{\pi}{4}\frac{f^2h_c(f)^2}{\rho_{\rm{c}}},
\end{equation}
where $f$ is the observed GW-frequency, $\rho_{\rm{c}}=3H^2/8\pi G$ is the energy-density required for a flat Universe, and $h_c(f)$ is the characteristic strain of the GW-background in a frequency interval centred at $f$.

The characteristic strain spectrum of a GW-background resulting from inspiraling SMBHBs is approximately $h_c(f)\propto f^{-2/3}$, and to a good approximation the strain-spectrum of a background from other sources can also be described by a power-law. Some measurable primordial background contributions may have a power-law index of $-1$ \citep{grishchuk-1976,grishchuk-2005}, while the background from decaying cosmic strings \citep{vilenkin-1981a,vilenkin-1981b,olmez-2010,sanidas-2012} may have $-7/6$ \citep{damour-vilenkin-2005}. For most models of interest, we can describe the strain-spectrum of a stochastic GW-background by $h_c(f) = A\left(f/\rm{yr}^{-1}\right)^{\alpha}$ \citep{jenet-2006}. This characteristic strain-spectrum is related to the one-sided power spectral density of the GWB-induced TOA-deviations by,
\begin{equation}\label{eq:gwb_power_spec}
S(f) = \frac{1}{12\pi^2}\frac{1}{f^3}h_c(f)^2 = \frac{A^2}{12\pi^2}\left(\frac{f}{\rm{yr}^{-1}}\right)^{-\gamma}\rm{yr}^3,
\end{equation} 
where $\gamma\equiv3-2\alpha$.

If a GWB is produced by processes occurring in the early Universe, or a superposition of high-redshift signals, then one would expect the resulting background to be reasonably isotropic. However, if the background signal is dominated by close, bright sources over the entire PTA-sensitivity band (as studied by \citet{ravi-2012}), or even just at high frequencies, the breakdown of stochasticity and resulting deviation from isotropy motivates a more general characterisation of the angular distribution of GW-power on the sky.

Over the last several years constraints on the amplitude of an \textit{isotropic} GWB have been published by the three major PTA collaborations \citep{demorest-2012,yardley-2011,van-haasteren-limits-2011}, the European Pulsar Timing Array (EPTA) \citep{epta-site}, the North American Nanohertz Observatory for Gravitational Waves (NANOGrav) \citep{nanograv-site}, and the Parkes Pulsar Timing Array (PPTA) \citep{ppta-site}. The International Pulsar Timing Array (IPTA) \citep{ipta-site} consortium combines these three efforts. The techniques used by these groups rely on the fact that a GWB induces correlated TOA-deviations between separated pulsars, which provides leverage against the undesirable noise processes. In fact, for an isotropic GWB this cross-correlation has a unique, distinctive angular signature dependant only on the angular-separation of the pulsars, and is commonly referred to as the Hellings and Downs curve \citep{hellings-downs-1983}.

Recently, \citet{cornish-sesana-2013} proposed a cross-correlation statistic aimed at targeting the brightest sources in each PTA frequency-band to account for anisotropy in the signal power. This statistic modifies the existing isotropic correlation techniques to account for finiteness of the source population. Essentially, instead of integrating the directional correlation function over the sky to produce the Hellings and Downs curve, this modified statistic sums the cross-correlation function over the brightest sources to infer their sky-positions. 

Alternatively, \citet{chiara-paper} have generalised the existing isotropic cross-correlation technique to account for arbitrary levels of anisotropy in a GWB. This is achieved by decomposing the angular power distribution in terms of spherical harmonics. Isotropy is represented by the monopole solution. For higher multipoles, the cross-correlation is no longer a simple function of pulsar angular separation, but rather depends on the position of each pulsar relative to orientation of the background anisotropy. Thus the concept of a one-parameter relationship to describe correlations induced by a GWB is no longer appropriate, and the relationship will instead become PTA- and GWB-specific, depending on the observed pulsars and the amount of anisotropy in the GWB.

In this paper we develop and test the first anisotropic Bayesian search-pipeline for the characterisation of a GWB using PTAs, employing the previously mentioned generalised spherical-harmonic decomposition of the cross-correlation. We have developed a suite of new plugins for the popular pulsar-timing analysis package \textsc{Tempo2} \citep{tempo2-1,tempo2-2,tempo2-3}, which can be used to inject GWBs with any user-defined anisotropy. For this first investigation of measuring the anisotropic-correlations, we inject backgrounds composed of typically $10^4$ sources, with sources placed according to various tested angular distributions. As such, we are investigating continuous anisotropy in the distribution of source-positions, rather than anisotropy through finiteness of the background, which we defer to future work.

This paper is organised as follows. In Section \ref{sec:ExpectedAnisSection} we perform some theoretical estimates of the expected level of anisotropy in a background composed of a superposition of many GW-signals, where the anisotropy either derives from varying source brightness or sparseness of the source-population. In Section \ref{sec:GW_Correlations} we review the theory of time- and angular-correlations of GWB-induced timing-residuals in pulsar signals, and outline a generalised correlation formalism for modelling correlations in anisotropic backgrounds. The theory of Bayesian inference is briefly reviewed in Sec.\ \ref{sec:bayesian-inference}, followed by a description of the data-analysis stages involved in the search for a stochastic GWB in pulsar datasets. A pipeline for producing simulated pulsar datasets containing an anisotropic GWB signal is described in Sec.\ \ref{sec:AnisPlugins}, followed by the results of Bayesian parameter-estimation and evidence-evaluation on these simulated datasets in Sec.\ \ref{sec:results}, with our conclusions in Sec.\ \ref{sec:conclusions}.


\section{The expected level of anisotropy}\label{sec:ExpectedAnisSection}

\begin{figure*}
   \subfloat[]{\incgraph{0}{0.5}{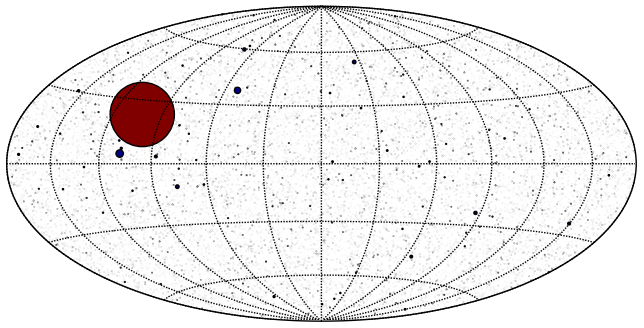}} 
   \subfloat[]{\incgraph{0}{0.5}{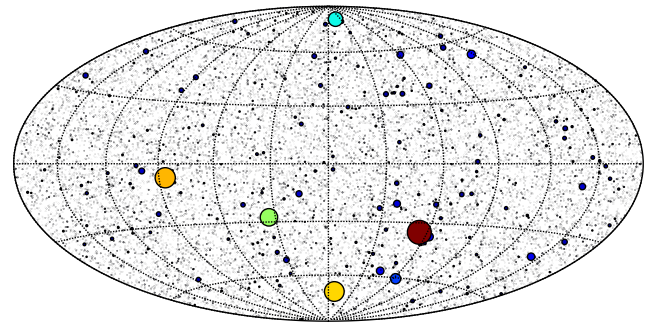}}
   \caption{\label{fig:Alberto_Datasets}Skymaps of GW source-populations generated by evolving a population of SMBHB systems. There are $\sim 2\times 10^4$ systems in each catalogue, which are typically massive ($10^7 - 10^{10} M_{\odot}$) and close ($z<2$). The relative size and colour of points within each skymap is indicative of the GW energy-flux from each system. The GW signal from the first dataset in (a) is clearly dominated by one very bright source. In the second dataset (b) we have several bright sources, however no outliers as in the first dataset.} 
 \end{figure*}

\begin{figure*}
   \subfloat[]{\incgraph{0}{0.5}{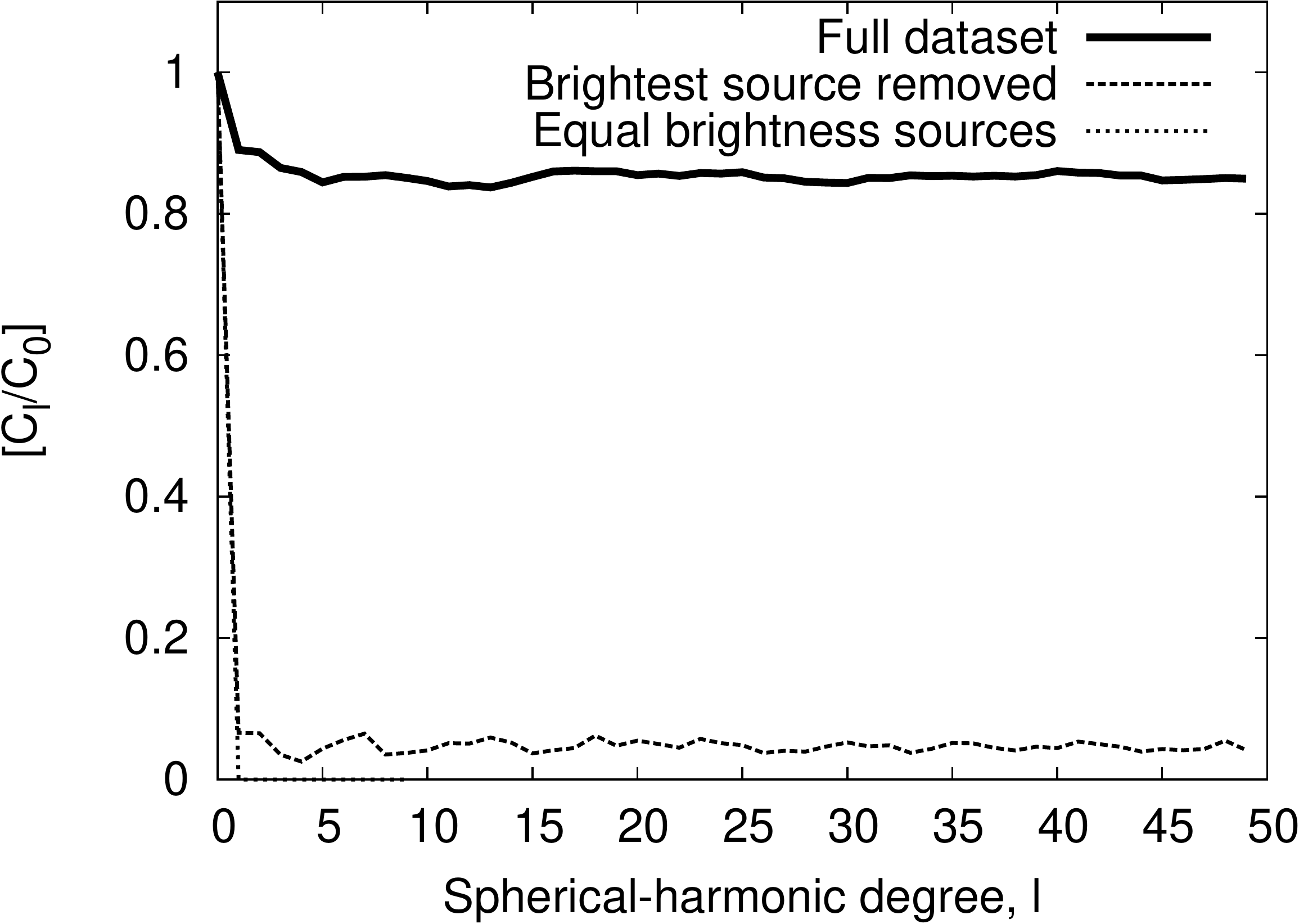}} 
   \subfloat[]{\incgraph{0}{0.5}{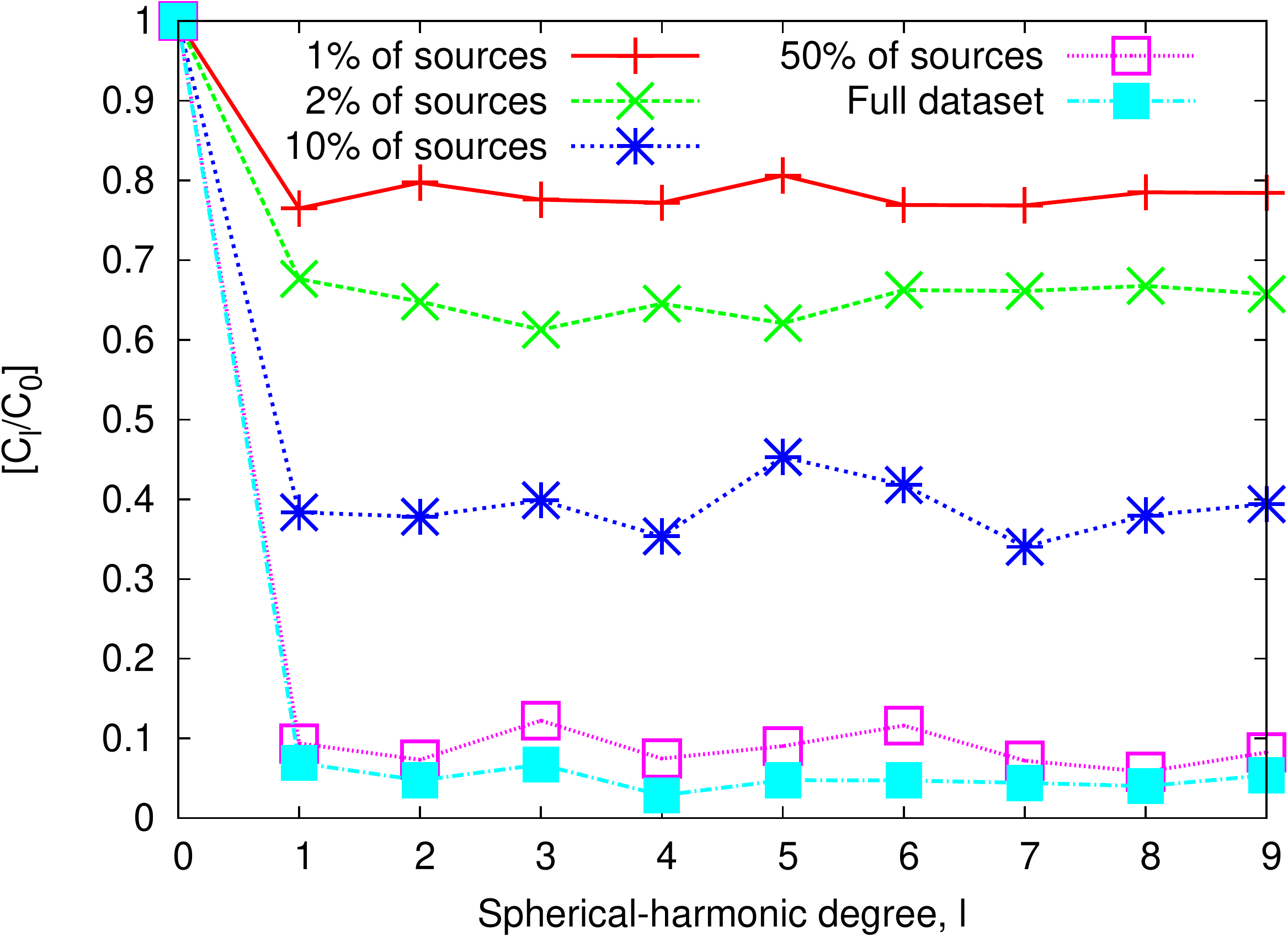}}
   \caption{\label{fig:Angular_Power_Spectrum}We perform a spherical-harmonic decomposition of the distribution of the GW energy density on the sky resulting from a population of inspiraling SMBHB systems, and compute the angular power spectrum, $C_l$. In (a) the influence of one very bright source in skymap-(a) of Fig.\ \ref{fig:Alberto_Datasets} clearly swamps all other influences, such that we almost trivially satisfy Uns\"{o}ld's theorem. If we remove the brightest source, then the anisotropic power drops to less than $0.1$ relative to the monopole (isotropic). Even this level of anisotropy is most likely due to the second-brightest sources, since if we assume equal brightness sources the $l\neq 0$ power drops to $\sim 10^{-5}$. This is expected for a large population of isotropically distributed source-positions. In (b) we analyse skymap-(b) of Fig.\ \ref{fig:Alberto_Datasets}. The full dataset is predominantly isotropic, however if we simulate source sparseness or intrinsic anisotropy by analysing sub-populations, then the level of anisotropy can be quite high.} 
 \end{figure*}

We wish to motivate a Bayesian search-pipeline which is generalised to arbitrary levels of anisotropy in the power distribution of the GW background. We take a realistic population of SMBHBs, generated by grafting baryonic physics onto the dark-matter haloes of a Millennium Simulation realisation, and extract those systems with observed GW frequencies which fall within PTA observation-frequencies \citep{sesana_systematic_signal}. These are typically massive and close systems ($z<2$). A typical catalogue contains $\sim 2\times 10^4$ sources, each with associated chirp masses, $\mathcal{M}$, observed GW-frequencies, $f$, and redshifts, $z$. Most importantly, the source positions are distributed isotropically on the sky. Hence, with this many sources making up our background we would expect any anisotropy to derive from varying source brightnesses. The radiated power in GWs of each binary in the source rest-frame is estimated using the quadrupole formula,
\begin{equation}
\frac{dE}{dt} = \frac{32}{5}\left[\pi\mathcal{M}f(1+z)\right]^{10/3},
\end{equation}
where $G=c=1$. The radiated energy flux in the observer frame is then given by,
\begin{equation}
\mathcal{F} = \frac{dE}{dt}\times\frac{1}{4\pi D_L(z)^2},
\end{equation}
where $D_L$ is the source luminosity distance. The energy-density in GWs, $\rho$, received from each source is then simply given by $\mathcal{F}/c$.

We now perform a spherical-harmonic decomposition of the energy density from a GW-population, for the purposes of evaluating the angular power-spectrum. Thus,
\begin{equation}
\rho(\hat\Omega) = \sum^{\infty}_{l=0}\sum^{l}_{m=-l}c_{lm}Y_{lm}(\hat\Omega)
\end{equation}
from which we compute the anisotropy coefficients,
\begin{equation}
c_{lm} = \int_{S^2}d\hat\Omega\; \rho(\hat\Omega)Y_{lm}(\hat\Omega)
\end{equation}
We define the observed angular power-spectrum as $C_l = \sum_m |c_{lm}|^2/(2l+1)$. Since we are dealing with a population of point sources, the energy-density distribution is just a sequence of delta-functions at the source positions, such that the spherical-harmonic decomposition becomes
\begin{equation}
c_{lm} = \sum_{i=1}^N\; \rho_iY_{lm}(\hat\Omega_i)
\end{equation}
which is just a summation over the spherical-harmonic functions evaluated at the $N$ source-positions, and weighted by the GW energy-density of each source.

We analyse two catalogues, whose skymaps are shown in Fig.\ \ref{fig:Alberto_Datasets}. The relative size and colour of each source is indicative of the GW energy-density, where larger and redder denotes a brighter source. The first dataset clearly includes a source which swamps the signal, while the second dataset is dominated by several bright sources. Figure \ref{fig:Angular_Power_Spectrum}(a) shows the angular power-spectrum of the first dataset, evaluated up to $l=50$. We see that the single brightest source swamps the power at all $l$, such that we are almost trivially satisfying Uns\"{o}ld's Theorem by having $[C_l/C_0] = 1$ at each $l$. If we remove the brightest source, then we get a power-spectrum in which the power is predominantly in the monopole (isotropic), with typically less than $10\%$ of the isotropic power in the other $l$ values. This small level of power in other $l$ values most likely derives from the few next-brightest sources, since if we fix all source brightness to the same value we obtain a power fraction of $\sim 10^{-5}$ in $l\neq 0$ values, which is what we expect when analysing a large population of equal-brightness sources which have been isotropically distributed on the sky. 

The second dataset has several very bright sources, but no outliers like in the first dataset. When analysed, most of the power lies in the monopole, with typically less than $10\%$ of the isotropic power in modes with $l\neq 0$. However, if we simulate source sparseness, or some form of intrinsic anisotropy through source-clustering, then the potential level of anisotropy is much higher. In Figure \ref{fig:Angular_Power_Spectrum}(b) the dataset has been shuffled and split into sub-populations. We compute the angular power-spectrum of each sub-population, and display the $95\%$ upper limit of the power in each $l$ derived from the ensemble of sub-populations. With a very sparse population the power in the modes with $l\neq 0$ could conceivably be $\lesssim 80\%$ of the isotropic power. This anisotropic power fraction tends to the value for the full dataset as we include more sources in the sub-populations. 

Thus, if we have several very bright sources dominating our background, or some form of intrinsic anisotropy through source-clustering, then the level of anisotropy in the sky-distribution of GW energy-density could conceivably be high. In the following, we simulate anisotropic backgrounds through a continuous anisotropy in GW source-positions, and test the robustness of a Bayesian anisotropic search-pipeline.

\section{Correlations induced by a GWB}\label{sec:GW_Correlations}

In this section we provide a brief overview of the time- and angular-correlations in the GW-induced deviations to the TOA of pulsar signals. We reiterate some of the formalism of \citet{chiara-paper} for ease of reference. 

\subsection{Cross-correlating the timing-residuals}

The redshift of a signal from a pulsar in the direction of unit vector $\hat{p}$, induced by the passage of a GW propagating in the direction of $\hat{\Omega}$ is \citep{anholm-2009,brook-flanagan-2011},
\begin{equation}\label{eq:pulse-freq-shift}
z(t,\hat\Omega) = \frac{1}{2}\frac{\hat{p}^a\hat{p}^b}{1+\hat\Omega\cdot\hat{p}}\Delta h_{ab}(t,\hat\Omega),
\end{equation}
where $\Delta h_{ab}\equiv h_{ab}(t_e,\hat\Omega)-h_{ab}(t_p,\hat\Omega)$, is the difference in the metric perturbation at the solar system barycentre, $h_{ab}(t_e,\hat\Omega)$, and at the pulsar, $h_{ab}(t_p,\hat\Omega)$. The vectors $(t_e,\hat{x}_e)$ and $(t_p,\hat{x}_p)$ give the spacetime coordinates of the solar system barycentre and pulsar, respectively.

This frequency-shift is \jon{then}{} integrated over time to give the induced timing residuals, which describe the perturbation to the TOA of pulses from a given pulsar,
\begin{equation}\label{eq:induced-residuals}
r(t)\equiv\int_0^tz(t')dt'.
\end{equation}

In the transverse-traceless (TT) gauge the perturbation of the metric caused by a mass quadrupole-moment can be described in terms of the linear superposition of the amplitudes of the two GW-polarisations permitted in GR. Hence, the GR metric perturbation can be written as,
\begin{equation}
h_{ab}(t,\hat{\Omega}) = h_{+}(t,\hat{\Omega})e_{ab}^{+}(\hat{\Omega}) + h_{\times}(t,\hat{\Omega})e_{ab}^{\times}(\hat{\Omega})
\end{equation}
where $e_{ab}^{+,\times}(\hat{\Omega})$ are the polarisation tensors, defined in terms of orthonormal basis vectors around $\hat{\Omega}$,
\begin{align}
e_{ab}^{+}(\hat{\Omega}) &= \hat{m}_a\hat{m}_b - \hat{n}_a\hat{n}_b \nonumber\\
e_{ab}^{\times}(\hat{\Omega}) &= \hat{m}_a\hat{n}_b + \hat{n}_a\hat{m}_b.
\end{align}
In a spherical-polar coordinate system, where the direction of GW-propagation is $\hat{\Omega} = [\sin\theta\cos\phi,\sin\theta\sin\phi,\cos\theta]$, the polarisation basis vectors take the following form,
\begin{align}
\hat{m} &= [\sin\phi,-\cos\phi,0] \nonumber\\
\hat{n} &= [\cos\theta\cos\phi,\cos\theta\sin\phi,-\sin\theta].
\end{align}

For the purposes of the following, we can also express the GW metric perturbation at a position in our coordinate system, $\vec{x}$, in terms of a plane-wave expansion. In a co-ordinate system centred at the solar system barycentre, with the pulsar some distance $L$ away (such that $t_p = t_e - L = t - L$, $\hat{x}_e=0$, and $\hat{x}_p=L\hat{p}$) this has the form
\begin{equation}
h_{ab}(t,\vec{x}) \equiv \sum_{A=+,\times}\int_{-\infty}^{\infty}df\int_{S^2}d\hat{\Omega}\; h_{A}(f,\hat{\Omega})e^{2\pi if(t-\hat{\Omega}\cdot\vec{x})}e^{A}_{ab}(\hat{\Omega}).
\end{equation}
Using this, $\Delta h_{ab}(t,\hat\Omega)$ in Eq.\ (\ref{eq:pulse-freq-shift}) becomes,
\begin{align}
\Delta h_{ab}(t,\hat\Omega) =& \sum_{A=+,\times}\int_{-\infty}^{\infty}df\int_{S^2}d\hat{\Omega}\;e^{A}_{ab}(\hat{\Omega})h_{A}(f,\hat{\Omega})e^{2\pi ift} \nonumber \\
&\times\left[1-e^{-2\pi ifL(1+\hat{\Omega}\cdot\hat{p})}\right]
\end{align}
where $\hat{p}$ and $L$ are the position and distance to the pulsar, respectively. The redshift of a signal induced by the passage of a GW can now be written as,
\begin{align}
z(t) =& \sum_{A=+,\times}\int_{-\infty}^{\infty}df\int_{S^2}d\hat{\Omega}\; F^{A}(\hat{\Omega})h_{A}(f,\hat{\Omega})e^{2\pi ift} \nonumber\\
&\times\left[1-e^{-2\pi ifL(1+\hat{\Omega}\cdot\hat{p})}\right]
\end{align}
where the antenna beam pattern for each polarisation in the PTA formalism is,
\begin{equation}
F^{A}(\hat{\Omega}) \equiv \frac{1}{2}\frac{\hat{p}^a\hat{p}^b}{1+\hat\Omega\cdot\hat{p}}e^{A}_{ab}(\hat{\Omega}).
\end{equation}
GW-detection pipelines used in PTAs search for correlated perturbations induced by GWs affecting different pulsars. This influence is manifested through the perturbation to the expected TOA of a given pulse caused by the passage of the GWs, such that what we actually measure are a set of TOA-residuals. These residuals encode information about all unmodelled phenomena, such as detectors-noise, intrinsic pulsar spin-noise and pulse phase-jitter. Crucially though, these sorts of influences are uncorrelated between different pulsars. The residuals induced by GWs in different pulsars will be correlated, which provides us the leverage required to separate these from noise processes. We will briefly discuss the details of how raw TOAs are processed into TOA-residuals in Section \ref{sec:pulsar-timing-analysis}.

Explicitly evaluating Eq.\ (\ref{eq:induced-residuals}) we find,
\begin{align}
r(t) =& \frac{i}{2\pi}\sum_{A=+,\times}\int_{-\infty}^{\infty}df\int_{S^2}d\hat{\Omega}\; F^{A}(\hat{\Omega})\frac{h_{A}(f,\hat{\Omega})}{f}\nonumber\\
&\times\left[1-e^{-2\pi ifL(1+\hat{\Omega}\cdot\hat{p})}\right]\left[1-e^{2\pi ift}\right]
\end{align}
To determine the correlation function between the residual at time $t_0$ in pulsar $a$ with residual at time $(t_0 + t)$ in pulsar $b$, $\langle r_a^*(t_0)r_b(t_0+t) \rangle$, we replace the statistical ensemble average with a time average, where the maximum time over which we can average is the light-travel time to the pulsar. We remove the unobservable constant offset from $r(t)$ \citep{lee-massive-gravitons}. Furthermore, if we have a background of GWs composed of a superposition of individually unresolvable single-sources, and such a background is stationary, Gaussian and independently polarised, we have \citep{maggiore-2008,lee-non-einsteinian},
\begin{equation}
\langle h_{A}^*(f,\hat{\Omega})h_{A'}(f',\hat{\Omega}') \rangle = H(f)P(\hat{\Omega})\delta_{AA'}\delta(f-f')\delta^2(\hat{\Omega},\hat{\Omega}')
\end{equation}
where $H(f)$ describes the spectral content of the background, defined in terms of the characteristic strain-spectrum as $h_c(f)^2/(16\pi f)$. Also, $P(\hat{\Omega})$ is the angular distribution of GW-power on the sky, which we interpret here as the angular distribution of sources comprising the background.
Finally,
\begin{align}
\langle r_a^*(t_0)r_b(t_0+t) \rangle = \int_{0}^{\infty}df\; \Gamma_{ab}(f)\frac{H(f)}{f^2}e^{2\pi ift}
\end{align}
where,
\begin{align}\label{eq:gamma-ab-f}
\Gamma_{ab}(f) &\equiv \int_{S^2}d\hat{\Omega}\; P(\hat{\Omega})\kappa_{ab}(f,\hat{\Omega})\left[\sum_{A=+,\times}F_a^{A}(\hat{\Omega})F_b^{A}(\hat{\Omega})\right] \nonumber\\
\kappa_{ab}(f,\hat{\Omega}) &\equiv \left[1-e^{2\pi ifL_a(1+\hat{\Omega}\cdot\hat{p_a})}\right]\left[1-e^{-2\pi ifL_b(1+\hat{\Omega}\cdot\hat{p_b})}\right]
\end{align}
In this case $\Gamma_{ab}(f)$ takes on the role of the ``overlap-reduction function'' (ORF) due to non-co-located, non-aligned detectors, which is often seen in the LIGO/LISA literature. 

The one-sided power spectrum of the induced residuals is related to the auto-correlation function via the Wiener-Khinchin theorem,
\begin{equation}
S(f) = \int_0^{\infty}\langle r_a^*(t_0)r_b(t_0+t) \rangle e^{-2\pi ift} dt.
\end{equation}

\subsection{Angular-correlations induced by a GWB}\label{sec:angular-formalism}

We now shift attention to the cross-correlation $\Gamma_{ab}(f)$ between pulsars $a$ and $b$. If we have a GWB composed of a superposition of individually unresolvable single-sources, then we can express their angular distribution on the sky as a decomposition in spherical harmonics. For the remainder of this analysis we decompose in terms of the real spherical harmonics, which are a linear combination of their complex analogues.
\begin{equation}
P(\hat{\Omega}) \equiv \sum^{\infty}_{l=0}\sum^{l}_{m=-l}c_{lm}Y_{lm}(\hat{\Omega})
\end{equation}
Examining the form of $\Gamma_{ab}(f)$ in Eq.\ (\ref{eq:gamma-ab-f}), we see that the overlap reduction function can also be decomposed into a set of ``correlation basis-functions''.
\begin{equation}
\Gamma_{ab}(f) = \sum^{\infty}_{l=0}\sum^{l}_{m=-l}c_{lm}\Gamma^{(ab)}_{lm}(f)
\end{equation}
where,
\begin{equation}
\Gamma^{(ab)}_{lm}(f) \equiv \int_{S^2}d\hat{\Omega}\; Y_{lm}(\hat{\Omega})\kappa_{ab}(f,\hat{\Omega})\left[\sum_{A=+,\times}F_a^{A}(\hat{\Omega})F_b^{A}(\hat{\Omega})\right]
\end{equation}
are the generalised ORFs, or ``correlation basis-functions''.

Using the formalism of \citet{allen-ottewill-97}, we define a ``cosmic-frame'' and a ``computational-frame''. In the former, pulsars have their usual RA and DEC positions on the sky. The latter frame places one pulsar along the $z$-axis of the coordinate system, and the other pulsar in the $(x-z)$ plane. This choice permits simple analytic expressions for the $\Gamma^{(ab)}_{lm}(f)$ to be determined, but we must remember that each distinct pair of pulsars defines a unique computational-frame, so we must rotate the computed values of $\Gamma^{(ab)}_{lm}(f)$ back into the common cosmic-frame.

We now quote the computational-frame correlation basis-functions from \citet{chiara-paper}, which have been converted to their real-form. Unless otherwise stated, we assume that $L_a=L_b$ and we are in the large $fL$ limit, such that $\kappa_{ab}(f,\hat{\Omega})\sim 1+\delta_{ab}$ (where $\delta_{ab}$ is the Kronecker delta), and therefore $\Gamma_{ab}(f)$ is independent of $f$. Let the angular-separation between pulsars $a$ and $b$ be $\zeta$, and $\alpha=(1+\cos\zeta)$ and $\beta=(1-\cos\zeta)$.
\begin{equation}
\Gamma_{00} = N\frac{\sqrt{\pi}}{2}\left[1+\frac{\cos\zeta}{3}+4(1-\cos\zeta)\ln\left(\sin\frac{\zeta}{2}\right)\right]\kappa_{ab}
\end{equation}
The monopole ($l=m=0$) is the Hellings and Downs solution, as expected. We choose the normalisation to be $N=3/8\pi$ such that a pure isotropic background with $c_{00}=2\sqrt{\pi}$ will have $\Gamma_{aa}(f) = c_{00}\Gamma^{(aa)}_{00}(f) = 1$.

In the following we refer to $\Gamma_{lm}$ as $\Gamma_{l,m}$ to easily distinguish negative $m$ values. The dipole correlation basis-functions have the following form,
\begin{align}
\Gamma_{1,-1} &= 0 \nonumber\\
\Gamma_{1,0} &= -N\frac{1}{2}\sqrt{\frac{\pi}{3}}\left[\alpha+3\beta\left(\alpha+4\ln\left(\sin\frac{\zeta}{2}\right)\right)\right]\kappa_{ab} \nonumber\\
\Gamma_{1,1} &= N\frac{1}{2}\sqrt{\frac{\pi}{3}}\sin\zeta\ln\left[1+3\beta\left(1+\frac{4}{\alpha}\ln\left(\sin\frac{\zeta}{2}\right)\right)\right]\kappa_{ab}
\end{align}
An alternative phrasing of the correlation induced by a dipole anisotropy in the GWB was derived in \citet{anholm-2009}, but as shown in \citet{chiara-paper} it is completely consistent with the formalism above.
\begin{align}
\Gamma_{ab,\rm{dip}} =& N\pi(\cos\delta_a + \cos\delta_b)\left(\cos\zeta - \frac{4}{3} \right.\nonumber\\
&\left. - 4\tan^2\left(\frac{\zeta}{2}\right)\ln\left(\sin\frac{\zeta}{2}\right)\right)\kappa_{ab}
\end{align}
where $\delta_a$ and $\delta_b$ are the angular separations between pulsars $a,b$ and the direction of the dipole moment  of the GWB.

The quadrupole correlation basis-functions have the following form in the computational frame
\begin{align}
\Gamma_{2,-2} = &0 = \Gamma_{2,-1}\nonumber\\
\Gamma_{2,0} =\; &N\frac{1}{3}\sqrt{\frac{\pi}{5}}\left[\cos\zeta+\right. \nonumber\\
&\left.\frac{15\beta}{4}\left(\alpha(3+\cos\zeta)+8\ln\left(\sin\frac{\zeta}{2}\right)\right)\right]\kappa_{ab} \nonumber\\
\Gamma_{2,1} =\; &N\frac{1}{2}\sqrt{\frac{\pi}{15}}\sin\zeta\left[5\cos^2\zeta+15\cos\zeta\right. \nonumber\\
&\left.-21-60\frac{\beta}{\alpha}\ln\left(\sin\frac{\zeta}{2}\right)\right]\kappa_{ab} \nonumber\\
\Gamma_{2,2} = &-N\frac{1}{4}\sqrt{\frac{5\pi}{3}}\frac{\beta}{\alpha}\left[\alpha(\cos^2\zeta+4\cos\zeta-9)\right. \nonumber\\
&\left.-24\beta\ln\left(\sin\frac{\zeta}{2}\right)\right]\kappa_{ab}
\end{align}
The reason why the negative-$m$ $\Gamma_{lm}$ values are zero can be easily understood from the form of the negative-$m$ real spherical-harmonic functions. These are proportional to $\sin(|m|\phi)$, which is an odd function. Since the evaluation of $\Gamma_{lm}$ involves integrating over $\phi$ between $0$ and $\pi$, then this will trivially yield zero.

We must rotate these computational-frame coefficients back into the cosmic-frame by using the rotational transformations of the real-form spherical harmonics. Finally, for a prescribed array of pulsars with known positions, we can completely pre-compute the cosmic-frame correlation basis-functions. These can be read-in to a search pipeline to be multiplied with model-dependent anisotropy coefficients, $c_{lm}$. The $c_{lm}$ values will be the search parameters in such a pipeline.


\section{Bayesian inference}\label{sec:bayesian-inference}
Bayes' theorem states that the \textit{posterior} probability density function (PDF), $p(\vec{{\mu}}|D,\mathcal{H})$, of the parameters $\vec{{\mu}}$ describing a hypothesis model $\mathcal{H}$, and given data $D$ is
\begin{equation} \label{eq:bayes-theorem}
p(\vec{{\mu}}|D,\mathcal{H}) = \frac{p(D|\vec{{\mu}},\mathcal{H})p(\vec{{\mu}}|\mathcal{H})}{p(D|\mathcal{H})},
\end{equation}
where,
\begin{align}
p(D|\vec{{\mu}},\mathcal{H})&\equiv\mathcal{L}(\vec\mu)=\text{likelihood of data given parameters,}\nonumber\\
p(\vec{{\mu}}|\mathcal{H})&\equiv\pi(\vec\mu)=\text{prior PDF of parameters,}\nonumber\\
p(D|\mathcal{H})&=\mathcal{Z}=\text{Bayesian evidence.}
\end{align}

The Bayesian evidence, $\mathcal{Z}$, is the probability of the observed data given the model ${\cal H}$
\begin{equation}\label{eq:evidence}
\mathcal{Z} = \int \mathcal{L}(\vec\mu)\pi(\vec\mu)d^N\mu.
\end{equation}
For posterior inference within a model, ${\cal Z}$ plays the role of a normalisation constant and can be ignored. However, if we want to perform model selection then this evidence value becomes key. In Bayesian model comparison we compute the Bayes factor
\begin{equation}
\frac{p(\mathcal{H}_2|\vec D)}{p(\mathcal{H}_1|\vec D)} = \frac{p(\vec D|\mathcal{H}_2)p(\mathcal{H}_2)}{p(\vec D|\mathcal{H}_1)p(\mathcal{H}_1)}=\frac{\mathcal{Z}_2\times p(\mathcal{H}_2)}{\mathcal{Z}_1\times p(\mathcal{H}_1)}.
\end{equation}
where $p(\mathcal{H}_2)/p(\mathcal{H}_1)$ is the prior probability ratio for the two competing models. This can often be set to one, and we will do so in the remainder of this analysis. The Bayes factor is then just the evidence ratio. Since the evidence is the average of the likelihood over the prior volume, it automatically incorporates Occam's razor, which states that, all else being equal, a model with fewer parameters is favoured. Hypothesis ${\cal H}_1$ is chosen if the Bayes factor is sufficiently large. Jeffreys \citep{jeffreys1983} gave a scale interpretation for the Bayes factor, which is shown in Table \ref{tab:jeffreys-scale-bayes}.

\begin{table}
\caption{\label{tab:jeffreys-scale-bayes}An interpretation of the Bayes factor in determining which model is favoured, as given by Jeffreys \citep{jeffreys1983}.}
\begin{ruledtabular}
\begin{tabular}{c c l}
Bayes factor, $\mathcal{K}$ & $\ln(\mathcal{K})$ & Strength of evidence\\
\hline
$< 1:1$ & $< 0$ & Negative (supports $\mathcal{H}_1$)\\
$1:1$ to $3:1$ & $0 -1.1$ & Barely worth mentioning\\
$3:1$ to $10:1$ & $1.1 - 2.3$ & Substantial\\
$10:1$ to $30:1$ & $2.3 - 3.4$ & Strong\\
$30:1$ to $100:1$ & $3.4 - 4.6$ & Very strong\\
$> 100:1$ & $> 4.6$ & Decisive\\
\end{tabular}
\end{ruledtabular}
\end{table}

\section{Pulsar Timing Analysis}\label{sec:pulsar-timing-analysis}

Repeated observation and study of a pulsar leads to a catalogue of the arrival-times of its pulses. After the actual radio-telescope observations are performed, the data is processed into the form of pulsar parameter files (``.par'') and timing files (``.tim''). The parameter file contains first estimates of the pulsar timing-model parameters; these parameters describe deterministic contributions to the arrival times. The vector of measured arrival times will be composed of a deterministic and a stochastic contribution (from time-correlated stochastic signals which are modelled by a random Gaussian process),
\begin{equation}
\vec{t}^{\rm{arr}} = \vec{t}^{\rm{det}} + \delta \vec{t}^{\rm{rgp}}.
\end{equation}
The stochastic process is characterised by its auto-correlation,
\begin{equation}\label{eq:pre-fit-stoch-cov}
C_{ij}=\langle\delta t_i^{\rm{rgp}}\delta t_j^{\rm{rgp}}\rangle, 
\end{equation}
where the elements of the covariance matrix are parametrised by a set of parameters, $\vec\phi$. Using the Wiener-Khinchin theorem, we can then define the auto-correlation as the Fourier transform of the power spectral density,
\begin{equation}
C(\tau_{ij}) = \int_0^{\infty}S(f)\cos(f\tau_{ij})df,
\end{equation}
where $\tau_{ij}=2\pi\vert t_i - t_j \vert$, and $S(f)$ is the power spectral density of the time-series $\delta \vec{t}^{\rm{rgp}}$. An analytic expression for the auto-correlation of a time-series influenced by an underlying power-law PSD is given in \citet{van-haasteren-limits-2011}, and is used in the following.

We describe the stages of pulsar-timing analysis as either \textit{offline} or \textit{online}, depending on whether they are performed before or during the sampling of parameter-space via likelihood evaluations.

\subsection{Processing raw arrival-times \textit{(offline)}}
The ``.par'' and ``.tim'' files are fed to the \textsc{Tempo2} software package \citep{tempo2-1,tempo2-2,tempo2-3} which processes the raw arrival-times. A vector of ``pre-fit'' timing-residuals are computed using the first guesses, $\beta_{0,i}$, of the ``$m$'' timing-model parameters from the ``.par'' files. This first guess is usually precise enough to allow a linear approximation to be used in the TOA fitting procedure. The post-fit timing residual are then given by
\begin{equation}
\delta \vec{t} = \delta \vec{t}^{\rm{prf}} + M\vec\xi,
\label{eq:postfit}
\end{equation}
where $\delta \vec{t}^{\rm{prf}}$ are the pre-fit timing-residuals (length $n$), $\vec\xi$ is the vector of deviations from the pre-fit parameters (length $m$) defined as $\xi_a = \beta_a - \beta_{0,a}$, and $M$ is the $(n\times m)$ ``design-matrix'', describing how the residuals depend on the timing-model parameters. \textsc{Tempo2} does not take into account the possible time-correlated stochastic signal in the TOAs, but performs a weighted least-squares fit for the timing-model parameter values. Hence it is possible that some of the time-correlated stochastic signal is removed by this fitting procedure, which is undesirable.

The \textsc{Tempo2} analysis provides output-residuals and the design matrix, $M$. The design matrix describes the dependence of the timing residuals on the timing-model parameters. The output-residuals form the input data vector for further study. 

\subsection{The time-domain likelihood \textit{(online)}}
We now use the \textsc{Tempo2} output-residuals to search for the presence of any correlated stochastic signal affecting the pulse arrival times. We assume that the part of the stochastic signal removed by the fitting procedure is small, so that the \textsc{Tempo2} output-residuals are related linearly to the stochastic contribution to the residuals~\citep{van-haasteren-levin-2012} 
\begin{equation}
\delta \vec{t} = \delta \vec{t}^{\rm{rgp}} + M\vec\xi,
\end{equation}
where, in this case, $\delta \vec{t}$ refers to the output-residuals from \textsc{Tempo2}. We note that the $\vec\xi$ appearing in this equation is different from that appearing in Eq.~(\ref{eq:postfit}).

The stochastic timing residuals, $\delta \vec{t}^{\rm rgp}$, arise from a time-correlated stochastic process with covariance matrix $C$ (see Eq.~(\ref{eq:pre-fit-stoch-cov})). This covariance matrix may contain contributions from the GWB, white-noise from TOA-errors, and possibly red-timing noise which is uncorrelated between different pulsars. The likelihood of measuring post-fit residuals $\delta\vec{t}$, given stochastic parameters $\vec\phi$, and marginalising with flat-priors over all timing-model parameters is \citep{van-haasteren-levin-2012},
\begin{align}\label{eq:vh-marg}
\mathcal{L}(\delta\vec{t}\vert\vec\phi)&=\frac{1}{\sqrt{(2\pi)^{n-m}{\rm{det}}(G^{T}CG)}} \nonumber\\
&\quad\exp{\left(-\frac{1}{2}\delta\vec{t}^{T}G\left(G^{T}CG\right)^{-1}G^{T}\delta\vec{t}\right)},
\end{align}
where $G$ is the matrix constructed from the final $(n-m)$ columns of the matrix $U$ in the SVD of the design matrix, $M=U\Sigma V^*$. The matrix $G$ can be pre-computed and stored in memory for use in each likelihood calculation (hence the computation of $G$ is \textit{offline}). Equation (\ref{eq:vh-marg}), with appropriate priors on the stochastic parameters, not only provides a robust, unbiased Bayesian framework for the search for correlated signals in PTAs, but also incorporates a lossless data compression through the $G$-matrix projection \citep{van-haasteren-compression-2012}.

In the case of multiple pulsars, where we have processed them independently with \textsc{Tempo2}, the total covariance matrix, total $G$ matrix and total residual vector are given by,
\begin{equation}
C = \begin{pmatrix}C_{11} & C_{12} & \ldots\\C_{21} & C_{22} & \ldots\\\vdots & \vdots & \ddots\end{pmatrix},G = \begin{pmatrix}G_1 & & & \\ & G_2 & & \\ & & \ddots &\end{pmatrix},\delta\vec{t} = \begin{pmatrix}\delta\vec{t}_1 \\ \delta\vec{t}_2 \\ \vdots\end{pmatrix},\nonumber\\
\end{equation}
where $C_{ab}$ is the auto-covariance matrix between pulsars $a$ and $b$, $G_a$ are the individual pulsar timing-model marginalisation-matrices and $\delta\vec{t}_a$ are the individual pulsar residual vectors.  We can split $C_{ab}$ into contributions from various stochastic sources. In this paper we consider only the stochastic influence due to a GWB (which has correlation $\Gamma_{ab}$ between pulsar $a$ and $b$), TOA error-bars (white and uncorrelated between different pulsars), and intrinsic pulsar red-noise (uncorrelated between different pulsars). So, the covariance between the $i$th residual of pulsar $a$ and the $j$th residual of pulsar $b$ is,
\begin{equation}
C_{(ai)(bj)} = C^{\rm GW}_{(ai)(bj)} + C^{\rm TOA}_{(ai)(bj)} + C^{\rm RN}_{(ai)(bj)},
\end{equation}
where $C^{\rm TOA}_{ab}$, $C^{\rm RN}_{ab}$ and $C^{\rm GW}_{ab}$ are diagonal, block-diagonal and block-symmetric respectively.

The spectral-density of induced timing-residuals for the intrinsic red-noise is taken to have the same form as a GWB, with amplitude $A_{\rm{red}}$ and slope $\gamma_{\rm{red}}$. 

\section{Simulating an anisotropic background}\label{sec:AnisPlugins}

\begin{figure*}
   \subfloat[]{\incgraph{0}{0.5}{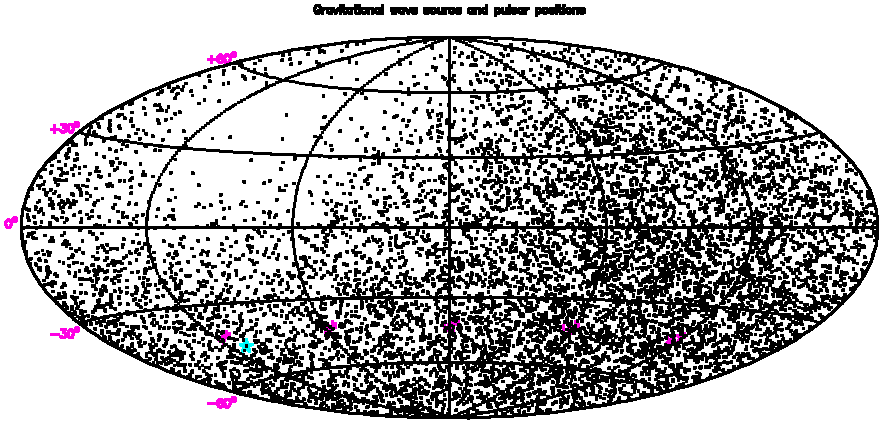}} 
   \subfloat[]{\incgraph{0}{0.5}{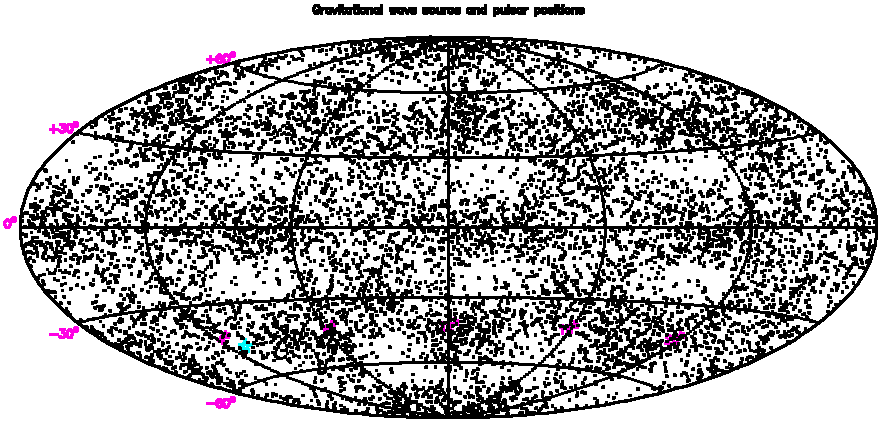}} \\
   \caption{\label{fig:tempo2-plugins}Screenshots from the \textsc{Tempo2} plugins. In (a) the over- and underdensity of sources in the different regions of the sky indicate the user-specified dipole direction using the {\bf GWdipolebkgrd} plugin. Likewise, the density of sources in (b) follows the user-specified multipolar structure of GW-power using the {\bf GWanisobkgrd} plugin.} 
 \end{figure*}

To simulate anisotropy in a GWB we have developed a suite of new \textsc{Tempo2} plugins. These permit a user to define not only the amplitude and spectral-index of the underlying characteristic strain spectrum of the background, as in the {\bf GWbkgrd} plugin, but also the angular distribution of the sources comprising the background. We can also define the polarisation of the sources comprising the background, which we have not limited to the Einsteinian polarisation states.

To generate an anisotropic background we define a probability density function for the placement of sources on the sky,
\begin{equation}\label{eq:probconstraint}
\mathcal{P}(\theta\in[\Theta,\Theta+d\theta],\phi\in[\Phi,\Phi+d\phi]) = P(\Theta,\Phi)d\theta d\phi
\end{equation}
where we enforce $P(\Theta,\Phi)\geq 0$. 

The background structure is user-defined, and is set by entering a multipolar decomposition of the sky-location PDF, as well as the usual dimensionless background-amplitude, $A$, and strain-spectrum slope, $\alpha$. We now give a brief description of the individual plugins, which will be made freely available.

\vspace{12pt}
{\bf GWdipolebkgrd} - Generates a background with a dipolar angular distribution of the GW-power. The user specifies the multipole coefficients, $\{c_{00},c_{1-1},c_{10},c_{11}\}$, or the direction of the dipole moment.

{\bf GWanisobkgrd} - Generates a background with arbitrary angular distribution of the GW-power. The user specifies the multipole coefficients, $\{ c_{lm}\}$.

{\bf GWgeneralbkgrd} - Generates an isotropic background composed of source-populations with different GW-polarisations, i.e.\ TT (GR modes), ST (scalar transverse, or {\it breathing}, mode), SL (scalar longitudinal) or VL (vector longitudinal). User specifies $A$ and $\alpha$ for each polarisation.

{\bf GWgeneralanisobkgrd} - Same as {\bf GWgeneralbkgrd}, but with an arbitrary angular distribution of sources for each polarisation, specified by a set of multipole coefficients, $\{ c_{lm}^A\}$, for each polarisation state $A$.
\vspace{12pt}

Figure \ref{fig:tempo2-plugins} shows screen-shots displaying the distribution of sources comprising several examples of user-specified backgrounds.

\section{Results}\label{sec:results}
We generated datasets similar to those used for Open Dataset 1 of the first IPTA Mock Data Challenge \citep{ipta-site}. They contained $36$ pulsars distributed across the sky and timed fortnightly for $5$ years. The injected signal was a GWB consistent with a background of inspiraling SMBHBs (i.e.\ $\gamma=13/3$), and the pulsars were timed to $100$ ns accuracy.

For a full dataset this amounts to $4680$ timing-residuals. Even with highly-tuned libraries and multithreading, the required matrix operations are very costly. We employ a high-fidelity data-compression technique developed by \citet{van-haasteren-compression-2012} for the purposes of throwing away unwanted high-frequency information and retaining as much information about the GWB as possible, which has a steep, red spectrum. 

The data-compression is based on diagonalising the expected signal covariance-matrix, and determining which basis vectors have the greatest contribution to the Fisher information. A set of reduced basis vectors are then found which capture as much information about the signal-of-interest as possible. In the lexicon of \citep{van-haasteren-compression-2012}, we demand a minimum fidelity of $0.99$. With $A=5\times 10^{-14}$, $\sigma_{\rm{WN}}=100$ ns and $0.99$ fidelity we can compress the number of residuals by approximately a factor of $5$. Matrix multiplications and inversions are $\mathcal{O}(n^3)$ operations, such that we achieve significant acceleration.


\subsection{Dipole injections}
\begin{figure}
   \incgraph{0}{0.5}{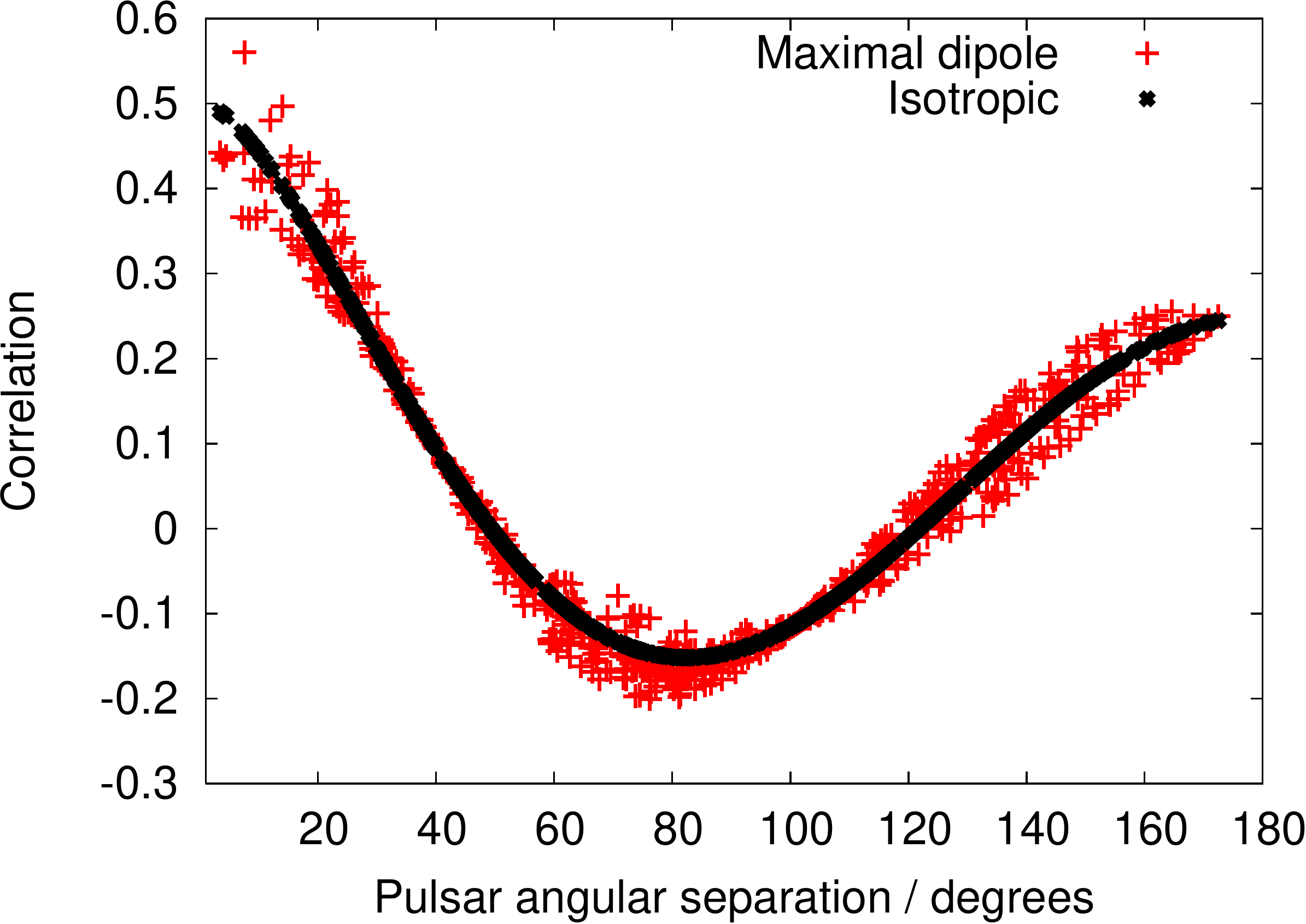}
   \caption{\label{fig:dipole-corr-display}Angular-correlations for all pulsar-pairings with non-zero angular-separation in the IPTA Mock Data Challenge array of $36$ pulsars. The black points indicate the angular-correlation when we have an isotropic distribution of GW-sources, which follows the expected Hellings and Downs curve. The red points show the angular-correlations when the angular-distribution of sources is proportional to $(1+\cos\theta)$, which corresponds to a dipole in the $+z$-direction. Crucially, the scatter of these points is not noise-related, but rather encodes information about the distribution of GW-sources with respect to the pulsar positions.}
 \end{figure}

The nature of a dipole anisotropy in a GW-source distribution can be seen in Fig.\ \ref{fig:tempo2-plugins}(a), where we see an over-density of sources in a particular sky-location, as well as a paucity of sources in the opposite sky-direction. Figure \ref{fig:dipole-corr-display} shows the angular-correlations of pulsar-pairings with non-zero angular-separation, $\Gamma_{ab}$, for all pulsars in our array. We also show the isotropic angular-correlation, which is the familiar Hellings and Downs curve \citep{hellings-downs-1983} and only depends on the angular-separation of the pulsars. However if we have a dipole anisotropy in the angular distribution of GW-sources on the sky, then the resultant pulsar angular-correlations depend not only on the angular-separation between the pulsars, but also on the absolute positions of the pulsars relative to the GW-source anisotropy. Hence pulsar-pairs which may have equal angular separations will not necessarily have equal angular-correlations if there is anisotropy in the GW-source population. Fig.\ \ref{fig:dipole-corr-display} shows the angular-correlation when the angular distribution of sources has the maximum dipole amplitude that satisfies Eq.~(\ref{eq:probconstraint}) and the dipole-direction is along the $+z$-axis.  Such a maximal dipole distribution is proportional to $(1+\cos\theta)$.

We see that a dipole anisotropy induces a small deviation away from the isotropic Hellings and Downs curve, which encodes information about the angular structure of the GWB.

We generated datasets with dipole anisotropies in various different directions on the sky, and used the formalism of Sec.\ \ref{sec:angular-formalism} to constrain the properties of the underlying GWB spectrum, as well as the direction of the dipole. For the latter, the alternative phrasing of dipole anisotropy of \citet{anholm-2009} is useful, since the dipole-correlation is simply expressed as a function of the pulsar-pulsar separations and the pulsar-dipole separations.

\begin{figure*}
   \subfloat[]{\incgraph{0}{0.5}{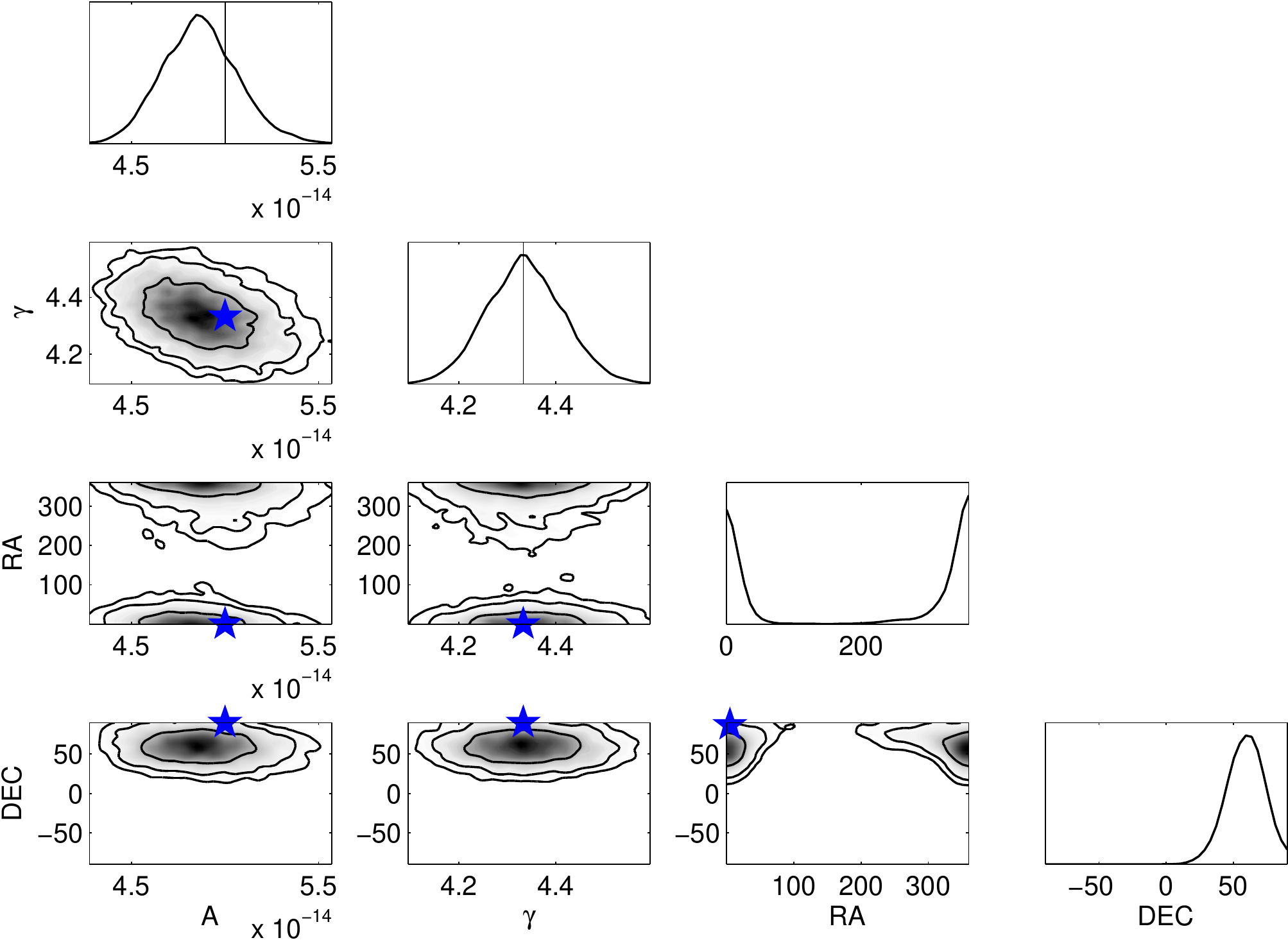}} 
   \subfloat[]{\incgraph{0}{0.5}{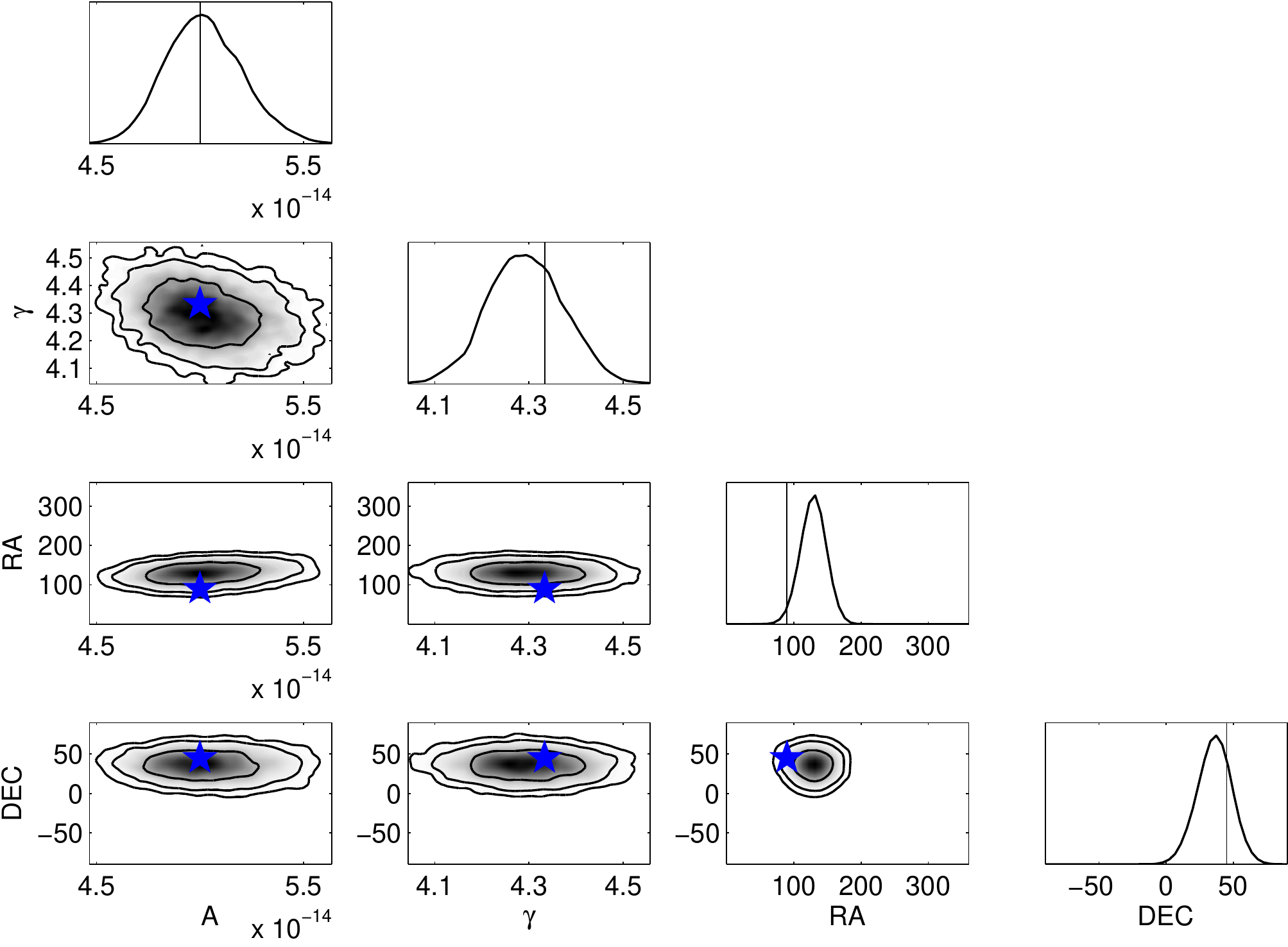}} \\
   \subfloat[]{\incgraph{0}{0.5}{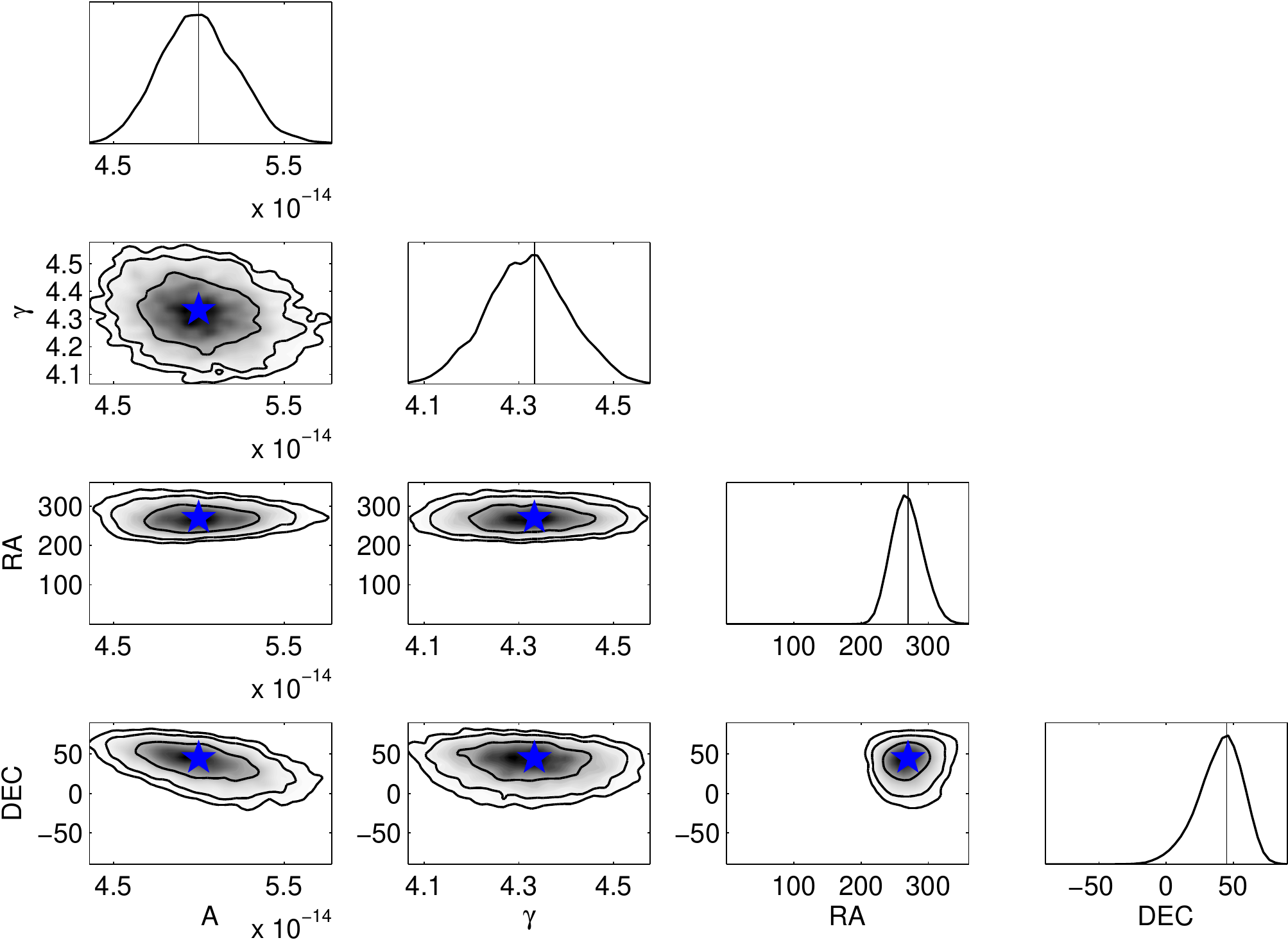}}
   \subfloat[]{\incgraph{0}{0.5}{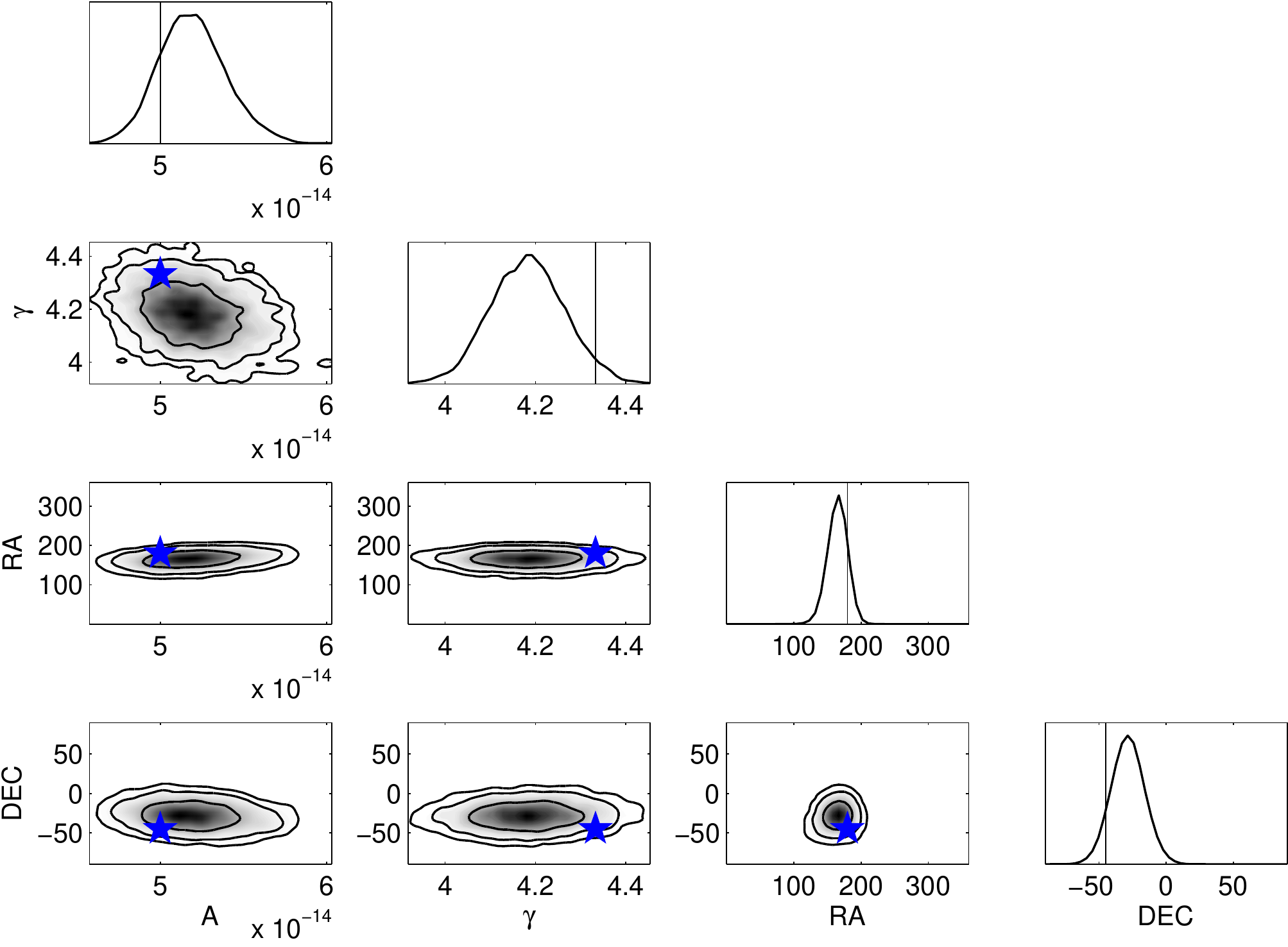}}
   \caption{\label{fig:DipDirectionRecon_4D}Recovered posterior probability distributions for a Bayesian parameter-estimation of four different dipolar GWBs in simulated PTA data. The anisotropy formalism is able to recover the direction of the injected dipole, as well as the properties of the underlying GWB spectrum. The strain-spectrum is a power-law in all cases, with $(A=5\times 10^{-14},\gamma=13/3)$, and the injected dipole-direction is (a) (RA$=0^{\circ}$,DEC$=90^{\circ}$), (b) (RA$=90^{\circ}$,DEC$=45^{\circ}$), (c) (RA$=270^{\circ}$,DEC$=45^{\circ}$), (d) (RA$=180^{\circ}$,DEC$=-45^{\circ}$). There is negligible correlation between the spectral-index $\gamma$ and the dipole-direction, which indicates that we can fix $\gamma$ to its injected value without biasing the anisotropy constraints. Fixing $\gamma$ affords the analysis large accelerations. We assume a maximal-dipole in the analysis of these datasets, i.e., the relative amplitude of the isotropic and dipole components is fixed to $1$ and not included as a model parameter.}
 \end{figure*}

The results of these injections is shown in Fig.\ \ref{fig:DipDirectionRecon_4D} and Fig.\ \ref{fig:DipDirectionRecon_3D}, where we show the recovered posterior probability distributions for the search-parameters. These posterior probability distributions show the ability of a Bayesian analysis to constrain the direction of a dipole in the GW-source population. Figure \ref{fig:DipDirectionRecon_4D} shows a search in which the spectral index, $\gamma$, was included as a parameter, whereas Fig.\ \ref{fig:DipDirectionRecon_3D} fixes $\gamma$ to the injected value of $13/3$. Fixing $\gamma$ accelerates the analysis since we can pre-compute various matrices and avoid some expensive matrix multiplications. As we can see from Fig.\ \ref{fig:DipDirectionRecon_4D} there is negligible correlation between $\gamma$ and the dipole-direction, so fixing $\gamma$ does not bias the reconstruction. 

The only inconsistency in the reconstructed direction was found when the dipole-direction was (RA=0.0, DEC=90.0). This arose because we sample uniformly in $\cos\theta$ so there is little prior-weight at high declinations. There is also the question of the significance of accounting for the anisotropy. In other words, does the anisotropic model provide a significantly better fit to the data than standard isotropic search would? We answer this question by inspecting the difference in $\ln(Z)$ between models which account for anisotropy, and for models which assume isotropy. The results are shown in Table \ref{tab:DipDirectionRecon_Bayes}, where we see that in all cases the true model is favoured, despite it having two extra dimensions compared to the isotropic model. However, we emphasise that the large Bayes' factors are partially the result of our injections being in the very strong-signal regime.

To test the robustness of the \textit{generalised} anisotropy formalism, we repeated the analysis of a maximal-dipole dataset in the direction (RA$=0^{\circ}$, DEC$=90^{\circ}$), but attempted a reconstruction of all the $c_{1,m}$ coefficients. We sample the anisotropy coefficients uniformly over the range $\in [-5,5]$. Since this is a strong anisotropic signal, and we expect to be likelihood-dominated, we adopt this uniform sampling range as a prior over the anisotropy coefficients here, and defer a proper discussion of a physically-motivated prior required for the analysis of real data until Sec.\ \ref{sec:Arb_Anis_Section}.

\begin{figure}
   \incgraph{0}{0.5}{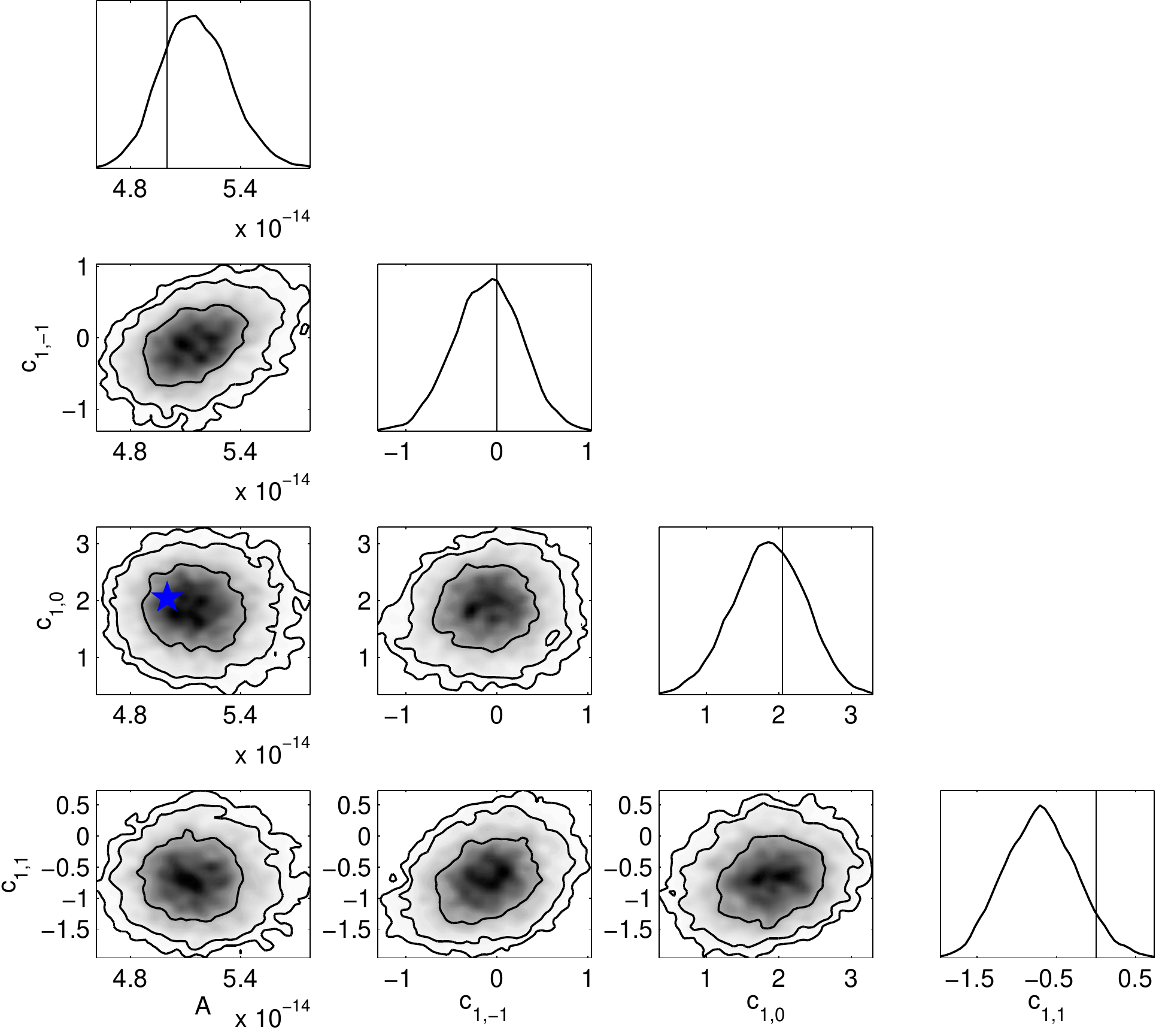}
   \caption{\label{fig:Dip_DipCorr}Results of a search over all dipole anisotropy coefficients, with the assumption relaxed that the relative amplitude of the dipole and isotropic components is maximal. The injected anisotropy was maximal-dipole with $c_{0,0}=2\sqrt{\pi}$ and $c_{1,0}=2\sqrt{\pi/3}$, corresponding to a dipole-direction of RA$=0.0^{\circ}$, DEC$=90^{\circ}$. The anisotropic-search pipeline fully recovered all details of the injected anisotropy, as well as the amplitude of the GWB characteristic strain-spectrum, $A$.}
 \end{figure}

We fix $c_{0,0}$ to $2\sqrt\pi$, and absorb the variation of the isotropic-power into the overall amplitude $A$. So, in this sense, we are searching for deviations from isotropy, rather than highly anisotropic backgrounds. Furthermore, by freely searching over $c_{1,m}$ we are removing our assumption of the anisotropy being maximal dipole. The results are shown in Fig.\ \ref{fig:Dip_DipCorr}, where we see that the recovered dipole-coefficients are completely consistent with the injected values of $\{c_{1,-1}=0,c_{1,0}=2\sqrt{\pi/3},c_{1,1}=0\}$, which correspond to a maximal-dipole in the $+z$-direction. The results of an analysis where we relax the assumption of \textit{dipole} anisotropy, and perform a search over all anisotropy coefficients up to and including quadrupole, are shown in Fig.\ \ref{fig:Dip_Arb_Anis}. The results, as before, are consistent with the injected anisotropy, however the increased dimensionality widens the posteriors such that we could conceivably be consistent with a large range of anisotropic distributions.

\begin{table}
\caption{\label{tab:DipDirectionRecon_Bayes}Evidence differences between an anisotropic and isotropic model when analysing datasets which have dipole-anisotropies injected in various different directions. In each case we favour the model which accounts for this anisotropy rather than the isotropic model. This is true even when the direction of the dipole is searched over, which adds an extra two dimensions to the parameter space. The Bayes' factor varies little when the spectral-index $\gamma\equiv 3-2\alpha$ is fixed to $13/3$, which is the expected value for a background composed of inspiraling SMBHBs.}
\begin{ruledtabular}
\begin{tabular}{c c c}
Dipole direction / degrees & \multicolumn{2}{c}{$\Delta\ln(Z)=\ln(Z_{\rm{true}})-\ln(Z_{\rm{iso}})$} \\
& $\gamma=13/3$ & $\gamma$ varied \\
\hline
RA=0.0 DEC=90.0 & 7.0 & 7.2 \\
RA=90.0 DEC=45.0 & 10.5 & 10.3 \\
RA=270.0 DEC=45.0 & 4.4 & 4.4 \\
RA=180.0 DEC=-45.0 & 7.3 & 7.5 \\
\end{tabular}
\end{ruledtabular}
\end{table}

We now investigate the degree to which an isotropic search is sub-optimal by generating many dipole-anisotropy datasets with different GWB amplitudes. The white-noise level is fixed at $100$ ns, such that we are effectively varying the signal-to-noise ratio (SNR) in the datasets. We parametrise SNR in the following way,
\begin{equation}
{\rm SNR} \sim \frac{\sigma_{\rm{GW}}}{\sigma_{\rm{WN}}}
\end{equation}
where,
\begin{align}\label{eq:sigmaGW}
\sigma_{\rm{GW}} &\sim 1.37\times\left(\frac{A}{1\times 10^{-15}}\right)\left(\frac{T}{\rm{yr}}\right)^{5/3}\;{\rm{ns,}}\quad{\rm{if}}\;\gamma=13/3\nonumber\\
\sigma_{\rm{WN}} &= 100\;{\rm{ns}}
\end{align}
and $\sigma_{\rm{GW}}$ is determined from an evaluation of the post-fit covariance function for power-law spectral densities. An estimate of this is obtained analytically by projecting the pre-fit covariance function into a new basis, which replicates the effect of fitting for quadratics. Given that the quadratic-spindown fitting has the dominant effect on the covariance matrix, Eq.\ (\ref{eq:sigmaGW}) provides a good estimate of $\sigma_{\rm{GW}}$ for given $A$ (and vice versa) \citep{lee-bassa-2012,van-haasteren-levin-2012,van-haasteren-compression-2012}.

\begin{figure}
   \incgraph{0}{0.5}{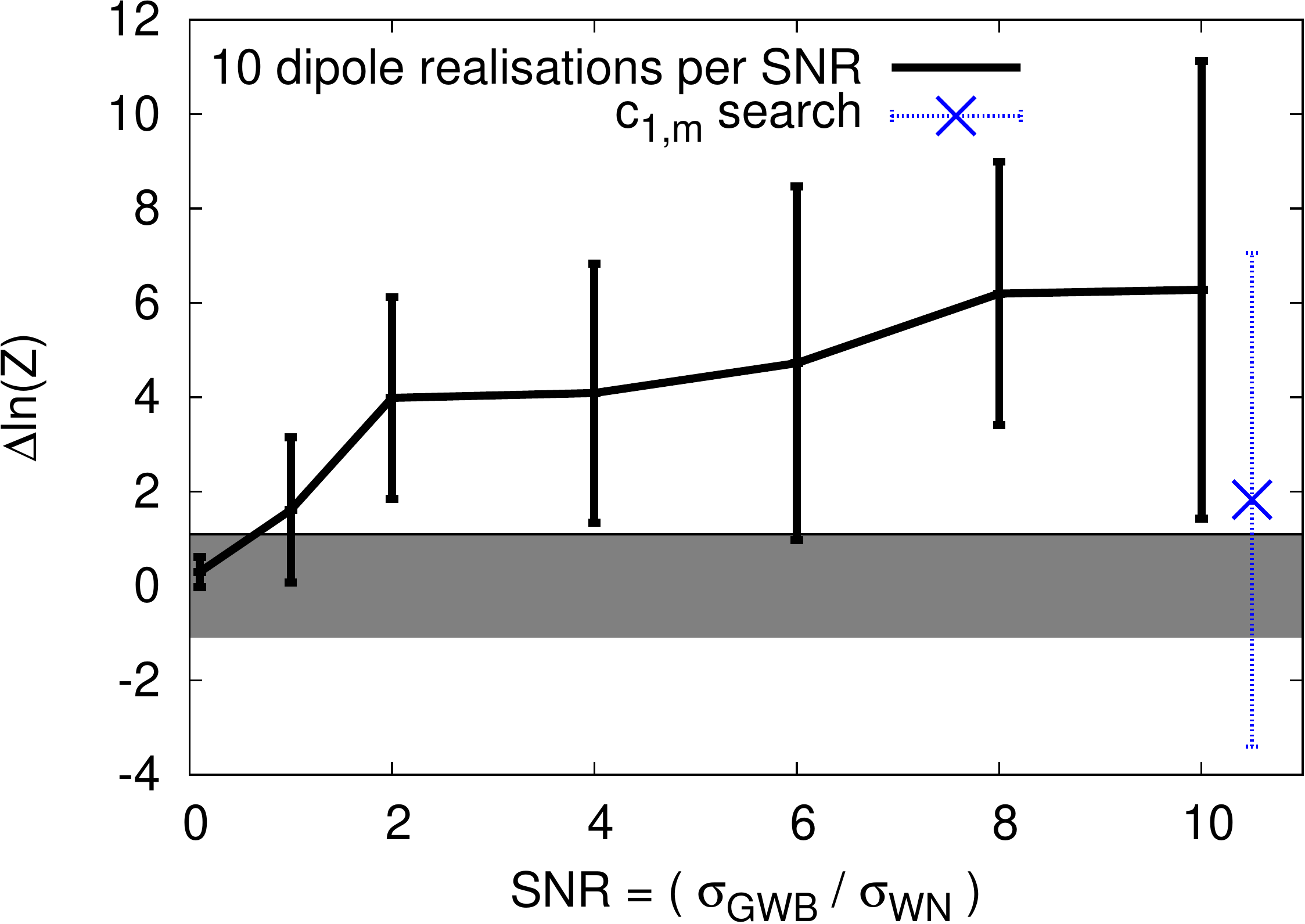}
   \caption{\label{fig:Dip_Bayes_SNR}Evidence for anisotropy versus isotropy as a function of the amplitude of the GWB. We generate many datasets with dipole-anisotropy such that the angular-distribution of GW-sources is $\propto (1+\cos\theta)$. We vary the SNR of the GWB by fixing the white-noise level and changing the amplitude of the characteristic strain-spectrum $A$, where for each SNR value we generate $10$ datasets. The SNR range $0.1-10$ corresponds to varying $A$ between $5\times 10^{-16} - 5\times 10^{-14}$. We fix $\gamma$ to its fiducial value of $13/3$, and evaluate the Bayesian evidence for models which assume the correct injected anisotropy ($c_{lm}$ fixed) and isotropy. The shaded region indicates the range of Bayes' factors for which we can not tell the difference between anisotropic and isotropic models. Since the injected noise is white-uncorrelated, and we compress to preserve low-frequency information, even at SNR $=0.1$ the GWB is loud enough to be detected with substantial evidence (see text for details). The blue, dotted point is a calibration-point for more general searches over $\{A,c_{1,m}\}$ at SNR $=10$, but offset on the $x$-axis for ease of viewing (see text for details).}
 \end{figure}

Varying the amplitude of the GWB between $5\times 10^{-16} - 5\times 10^{-14}$, gives an effective SNR span of $0.1-10$. At each SNR in this range we generated $10$ dataset realisations and evaluated the difference in $\ln(Z)$ values for an anisotropic and an isotropic model. This was done with the GWB correlations fixed to those corresponding to the injected level of anisotropy, which in this case was the maximal-dipole in the $+z$-direction described earlier. These results are shown in Fig.\ \ref{fig:Dip_Bayes_SNR}. We note that this figure does not show the true odds ratio for the anisotropic versus isotropic model, since the degrees of freedom corresponding to the dipole direction have not been included in the parameter space for the anisotropic model. Instead, this figure indicates a best case scenario for when the evidence would start to favour the anisotropic model. Also, one should note that because the injected noise is white-uncorrelated, and we have compressed to preserve information about a steep red-noise process, even at SNR $=0.1$ the GWB is loud enough to be detected with substantial evidence. At this SNR the largest noise-versus-(isotropic-)GWB log-Bayes factor is $0.03$, with a mean and median of $-6.30$ and $-2.62$, respectively. So, (in this parametrisation of SNR) at SNR $=0.1$ we can detect a GWB with angular-correlation of \textit{approximate} Hellings and Downs form, but the background is not yet loud enough for anisotropy-induced deviations to be visible.

We see that in these idealised circumstances the anisotropic model is marginally favoured for SNRs above $\sim 1$ and decisively favoured for SNRs above $\sim 5$. Including the dipole direction parameters (or searching over $c_{1,m}$) in the parameter space reduces the anisotropic evidence by several and so in practice SNRs of $\sim 5$--$10$ would be required before the presence of anisotropy would become clearly visible to an analysis of this type (see also Fig.~\ref{fig:RedBayesTrack_SNR} below). We provide a calibration point at SNR $=10$, where we search over $\{A,c_{1,m}\}$ for the $10$ dataset realisations at this SNR. We see that the increased dimensionality of the search penalises the recovered Bayes factors, however the average log-Bayes factor implies that the presence of anisotropy may become clearly visible at SNRs greater than $10$.

The datasets described so far have contained only white-noise. Now we generate datasets with an additional red-noise component uncorrelated between different pulsars. The red-noise shares a common spectrum, but with a different realisation for each pulsar. The spectral-index was taken to match that of the IPTA Open3 dataset, which was $\gamma_{\rm{red}}=1.7$. As before, we include an uncorrelated white-noise component of $100$ ns in each pulsar. Following Eq.\ (\ref{eq:sigmaGW}), we parametrise the SNR as,
\begin{equation}
{\rm SNR} \sim \frac{\sigma_{\rm{GW}}}{\sigma_{\rm{RED}}} \sim 0.422\times\left(\frac{A}{A_{\rm{red}}}\right)\times \left(\frac{T}{\rm{yr}}\right)^{79/60}
\end{equation}
where the form of $\sigma_{\rm{RED}}$ follows from the more general form of the rms-residual induced by a stochastic process with power-law spectrum of arbitrary slope \citep{van-haasteren-levin-2012}.

\begin{figure*}
   \incgraph{0}{1.0}{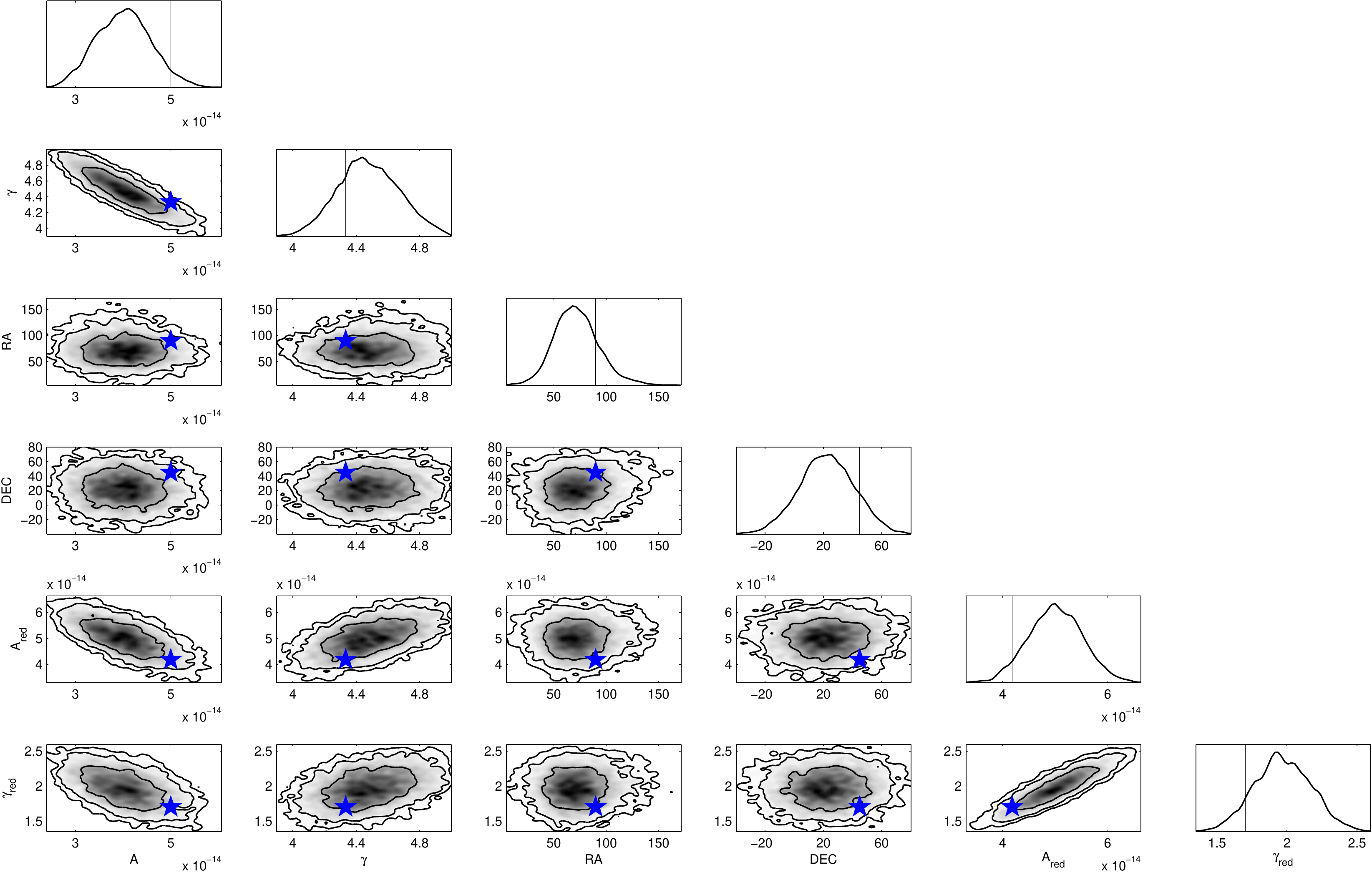}
   \caption{\label{fig:Dipole_RED_SNR10}An example of anisotropy reconstruction when uncorrelated red-noise is present in the dataset. The anisotropy is maximal-dipole, with direction (RA$=90^{\circ}$, DEC$=45^{\circ}$). The injected GWB parameters are ($A=5\times 10^{-14}$,$\gamma=13/3$). The red-noise is uncorrelated between different pulsars, but shares a common spectrum, with ($A_{\rm{red}}=4.2\times 10^{-14}$,$\gamma_{\rm{red}}=1.7$), such that we have an effective SNR of $\sim 4.2$. In this dataset, the $\ln(\mathcal{B})$ for anisotropic versus isotropic was $\sim 4.4$, which is borderline decisive evidence for anisotropy.}
 \end{figure*}

The analysis is more expensive than a GWB-only analysis, since we must also characterise and constrain the red-noise properties. Furthermore, the red-noise has a shallower spectrum, such that our residuals contain more high-frequency noise, and we can not compress to the same degree as in the case of a GWB-only dataset. Figure \ref{fig:Dipole_RED_SNR10} shows the results of an SNR $\sim 4$ analysis. We see that at this SNR we can characterise the GWB, red-noise and anisotropy properties. In fact, for this dataset realisation, the value of $\ln(\mathcal{B})$ for anisotropy versus isotropy was $\sim4.4$, which is borderline decisive evidence for anisotropy. Since the analysis is quite computationally expensive, and we wish to perform a more systematic study over many background realisations, we repeat this analysis but fix $\gamma$ and $\gamma_{\rm{red}}$ to the injected values of $13/3$ and $1.7$, respectively. Fixing these spectral indices accelerates the analysis, does not bias the parameter reconstruction (see Fig.\ \ref{fig:Dipole_RED_SNR10_GammaFixed}), and only raises the value of $\ln(\mathcal{B})=\Delta\ln(Z)$ by $\sim 0.8$.

We perform a similar study to Fig.\ \ref{fig:Dip_Bayes_SNR}, generating $5$ datasets per SNR, each with common GWB properties ($A=5\times 10^{-14}$,$\gamma=13/3$), $\gamma_{\rm{red}}=1.7$, but varying the red-noise amplitude, $A_{\rm{red}}$ to simulate the different SNRs. The injected anisotropy is maximal-dipole in the (RA$=90^{\circ}$, DEC$=45^{\circ}$) direction in all cases. 
We search over the parameters $\{A,A_{\rm{red}},$RA$,$DEC$\}$. The results are shown in Fig.\ \ref{fig:RedBayesTrack_SNR}, where we see that the presence of uncorrelated red-noise in the datasets can be accounted for, but its presence means that the SNR required to identify the presence of anisotropy in the data appears somewhat higher. However, in this case we include a search over the dipole direction in the evaluation of the evidence, so these results are much closer to what might be achievable in practice (although fixing the slope of the red noise spectrum in particular might not be a valid assumption). Fig.~\ref{fig:RedBayesTrack_SNR} suggests that even in the presence of pulsar red noise the evidence for anisotropy will become apparent for SNRs around $3$ and will become decisive for SNRs of $\sim 6$.

For the remainder of this analysis we ignore red-noise and inject only uncorrelated white-noise in the form of TOA error-bars, but expect that the effect of red noise will be to slightly increase the SNR (by a few) required for a confident detection of anisotropy.

\begin{figure}
   \incgraph{0}{0.5}{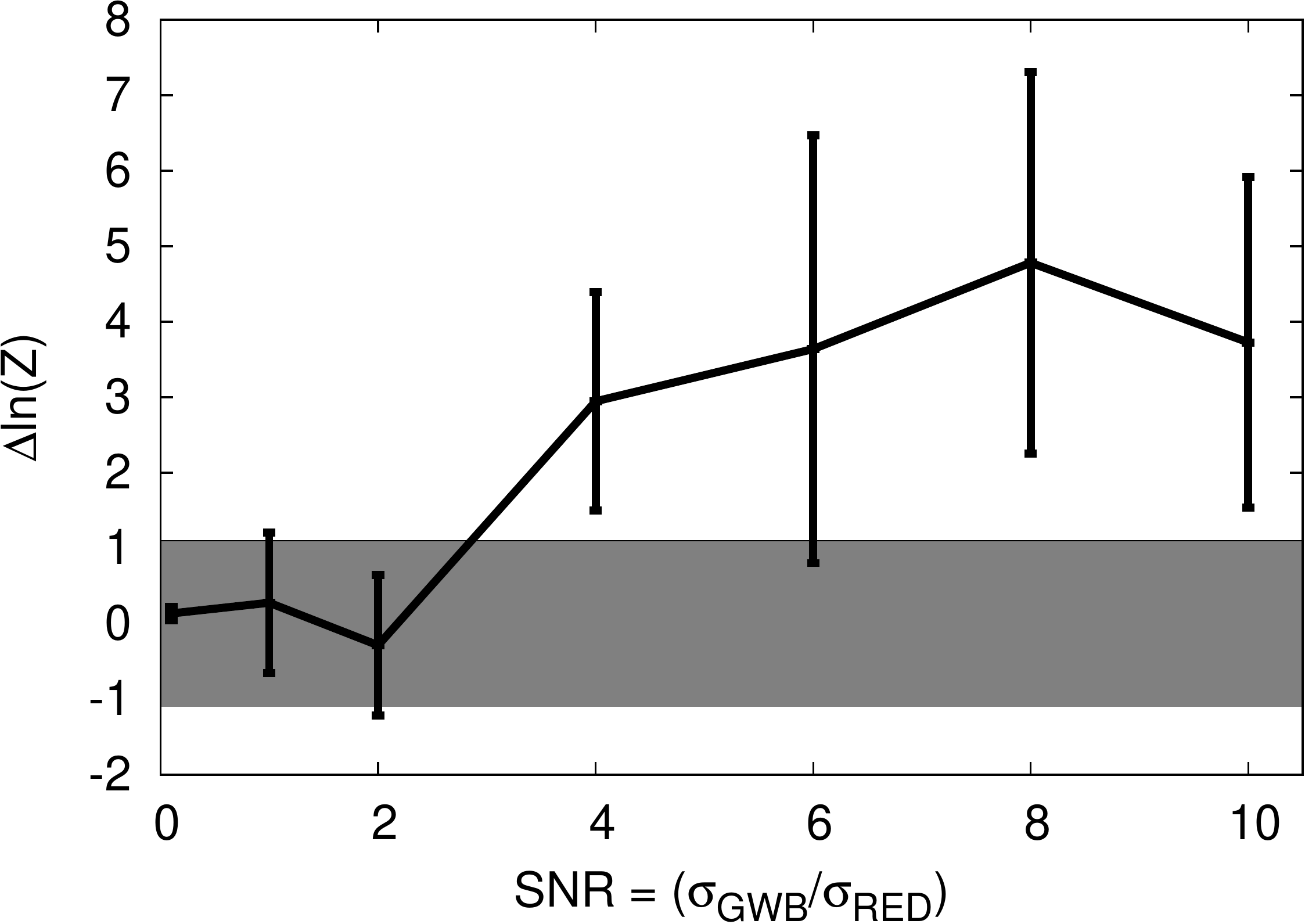}
   \caption{\label{fig:RedBayesTrack_SNR}As Fig.~\ref{fig:Dip_Bayes_SNR}, but now for datasets containing red noise. We generate many datasets with dipole-anisotropy such that the angular-distribution of GW-sources is $\propto 1+\cos\theta$. We vary the SNR of the GWB by fixing the characteristic strain-spectrum amplitude, $A$, at $5\times 10^{-14}$ and changing the amplitude of the red-noise level, where for each SNR value we generate $5$ datasets. The SNR range $0.1-10$ corresponds to varying $A_{\rm{red}}$ between $\sim 1.8\times 10^{-12} - 1.8\times 10^{-14}$. We fix the slopes of the power spectral-density of the induced residuals due to a GWB and uncorrelated red-noise to the injected values of $13/3$ and $1.7$, respectively, and evaluate the Bayesian evidence for models which assume a maximal-dipole anisotropy (direction searched over) and isotropy. The shaded region indicates the range of Bayes' factors for which we can not tell the difference between anisotropic and isotropic models.}
 \end{figure}

\subsection{Quadrupole injections}
\begin{figure}
   \incgraph{0}{0.5}{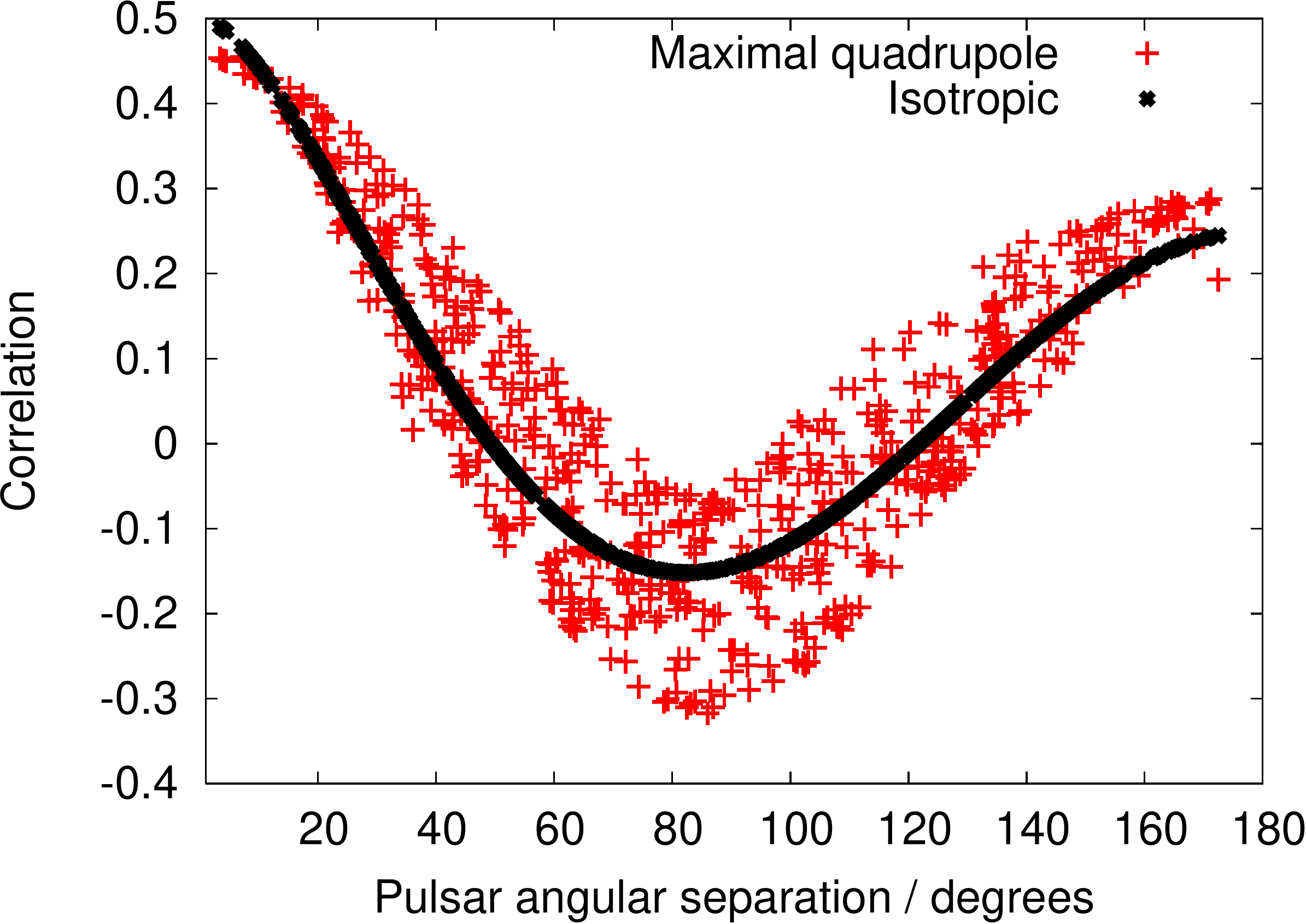}
   \caption{\label{fig:quadrupole-corr-display}Angular-correlations for all pulsar-pairings with non-zero angular-separation in the IPTA Mock Data Challenge array of $36$ pulsars. The black points indicate the angular-correlation when we have an isotropic distribution of GW-sources, which follows the expected Hellings and Downs curve. The red points show the angular-correlations when the angular-distribution of sources follows $\propto 1+(3\cos^2\theta-1)$, which corresponds to a quadrupole anisotropy. The quadrupole clearly induces a larger deviation from the Hellings and Downs curve than the dipole-anisotropy in Fig.\ \ref{fig:dipole-corr-display}.}
 \end{figure}

\begin{figure*}
   \subfloat[]{\incgraph{0}{0.5}{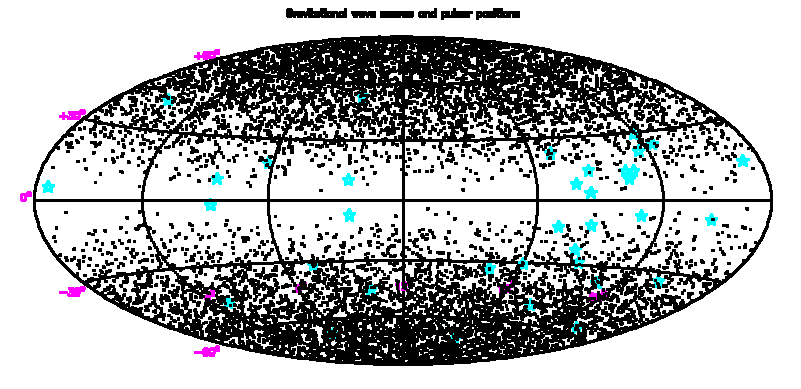}} 
   \subfloat[]{\incgraph{0}{0.5}{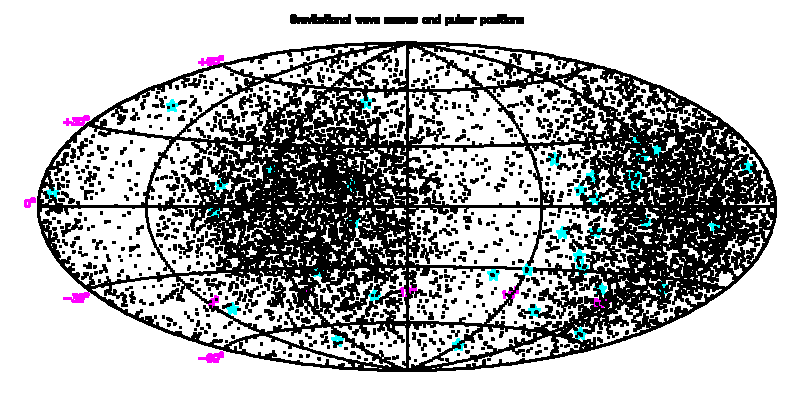}} \\
   \subfloat[]{\incgraph{0}{0.5}{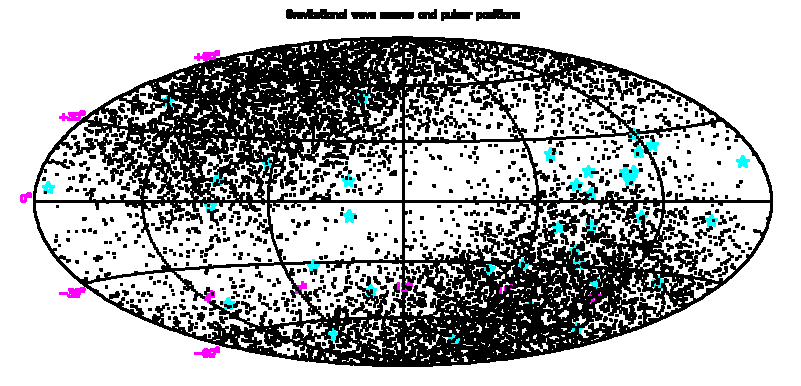}}
   \subfloat[]{\incgraph{0}{0.5}{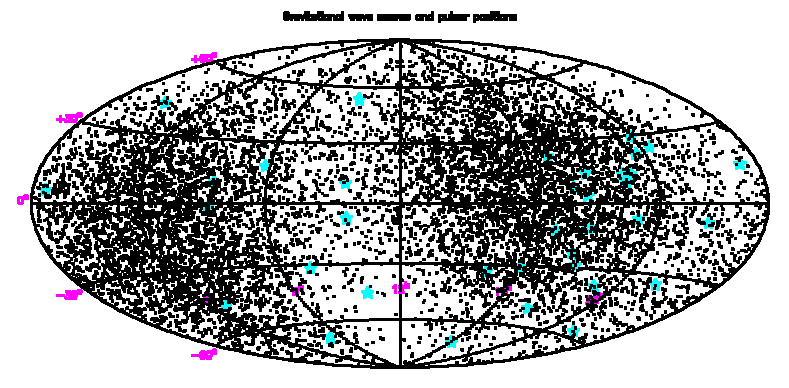}}
   \caption{\label{fig:GWanisbkgrd_Quad_Dists}Skymaps showing the quadrupolar angular-distribution of GW sources constituting various GWB realisations. The anisotropy-coefficients for each dataset are (a) $\{c_{0,0}=2\sqrt\pi;\;c_{2,0}=4\sqrt{\pi/5}\}$ (b) $\{c_{0,0}=2\sqrt\pi;\;c_{2,-2}=-0.9 c_{0,0};\;c_{2,0}=-0.3 c_{0,0}\}$ (c) $\{c_{0,0}=2\sqrt\pi;\;c_{2,-1}=-0.8 c_{0,0};\;c_{2,0}=0.5 c_{0,0};\;c_{2,2}=-0.3 c_{0,0}\}$ (d) $\{c_{0,0}=2\sqrt\pi;\;c_{2,-2}=0.56 c_{0,0};\;c_{2,-1}=0.20 c_{0,0};\;c_{2,0}=-0.32 c_{0,0};\;c_{2,1}=0.27 c_{0,0};\;c_{2,2}=-0.40 c_{0,0}\}$.} 
 \end{figure*}

The nature of a quadrupole-anisotropy in a GW-source distribution, with the maximal quadrupole angular-distribution allowed under the constraint~(\ref{eq:probconstraint}), which is $\propto 3\cos^2\theta$, can be seen in Fig.\ \ref{fig:GWanisbkgrd_Quad_Dists}(a). Figure \ref{fig:quadrupole-corr-display} shows the distinct angular-correlations for all pulsars in the array when there is a quadrupole anisotropy in the GW-source population. We see that a quadrupole-anisotropy, which in this case manifests itself through the angular-distribution of sources being $\propto 1+(3\cos^2\theta-1)$, induces a larger deviation of the correlations from the Hellings and Downs curve than the dipole. We might expect then that the assumption of isotropy will bias our results to a greater extent than in the dipole-anisotropy case.

We perform a similar analysis to Fig.\ \ref{fig:Dip_Bayes_SNR}, generating many datasets with fixed uncorrelated white-noise levels of $100$ ns, but varying the amplitude of the GWB to simulate different SNRs (see Eq.\ (\ref{eq:sigmaGW})). As in the dipole case, we vary the amplitude of the GWB between $5\times 10^{-16} - 5\times 10^{-14}$, giving an effective SNR span of $0.1-10$ and at each SNR we generate $10$ dataset realisations. The only parameter we search for in the analysis of each dataset is the amplitude, $A$, of the background. Hence, as before, this study represents a best case assessment of when we will be able to identify the presence of a quadrupole anisotropy (of the form $P(\theta,\phi)=3\cos^2\theta$) in the data. In reality, we will not know the direction of the quadrupole anisotropy in advance, which introduces extra dimensionality into the parameter space that must be searched over. This will tend to reduce the log-evidence value and so larger SNRs will be required to identify quadrupole anisotropies in practice. As in the dipole case, one should note that because the injected noise is white-uncorrelated, and we have compressed to preserve information about a steep red-noise process, even at SNR$=0.1$ the GWB is loud enough to be detected with substantial evidence. At this SNR the largest noise-versus-(isotropic-)GWB log-Bayes factor is $0.3$, with a mean and median of $-9.3$ and $-5.8$, respectively. As discussed previously, at SNR = $0.1$ the GWB is loud enough such that we can detect a background which has \textit{approximate} Hellings and Downs form, but not yet loud enough to allow inference of anisotropy-induced deviations.

\begin{figure}
   \incgraph{0}{0.5}{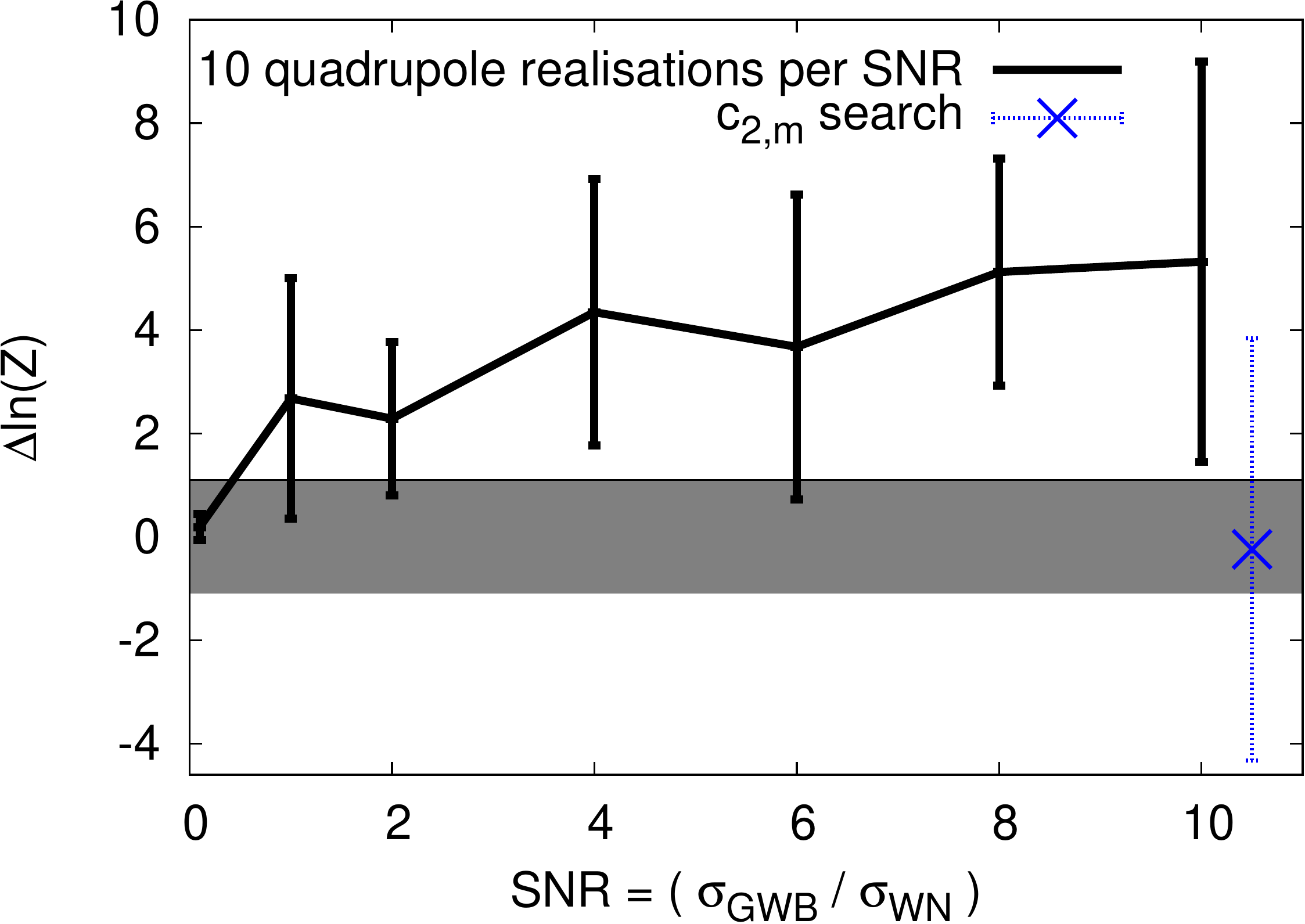}
   \caption{\label{fig:Quad_Bayes_SNR}As Fig.~\ref{fig:Dip_Bayes_SNR}, but now for datasets containing a maximal quadrupole-anisotropy such that the angular-distribution of GW-sources is $\propto 1+(3\cos^2\theta-1)$. We vary the SNR of the GWB by fixing the white-noise level and changing the amplitude of the characteristic strain-spectrum $A$, and for each SNR value we generate $10$ datasets. The SNR range $0.1-10$ corresponds to varying $A$ between $5\times 10^{-16} - 5\times 10^{-14}$. We fix $\gamma$ to its fiducial value of $13/3$, and evaluate the Bayesian evidence difference between a model that assumes the correct injected anisotropy ($c_{lm}$ fixed) and an isotropic model. The shaded region indicates the range of Bayes' factors for which we can not tell the difference between anisotropic and isotropic models. Since the injected noise is white-uncorrelated, and we compress to preserve low-frequency information, even at SNR $=0.1$ the GWB is loud enough to be detected with substantial evidence (see text for details). The blue, dotted point is a calibration-point for more general searches over $\{A,c_{2,m}\}$ at SNR $=10$, but offset on the $x$-axis for ease of viewing (see text for details).}
 \end{figure}

The results of this study can be seen in Fig.\ \ref{fig:Quad_Bayes_SNR}. These results are quite similar to the dipole case -- the presence of anisotropy becomes identifiable at an SNR of about 1 and the evidence is decisive for an SNR of $\sim 5$. We provide a calibration point at SNR $=10$, where we search over $\{A,c_{2,m}\}$ for the $10$ dataset realisations at this SNR. As expected, we see that the increased dimensionality of the search penalises the recovered Bayes factors, however the average log-Bayes factor implies that the presence of anisotropy may become visible with substantial evidence at SNR $=10$, but we will require SNR $>10$ for it to be clearly visible.

\begin{figure*}
   \subfloat[]{\incgraph{0}{0.5}{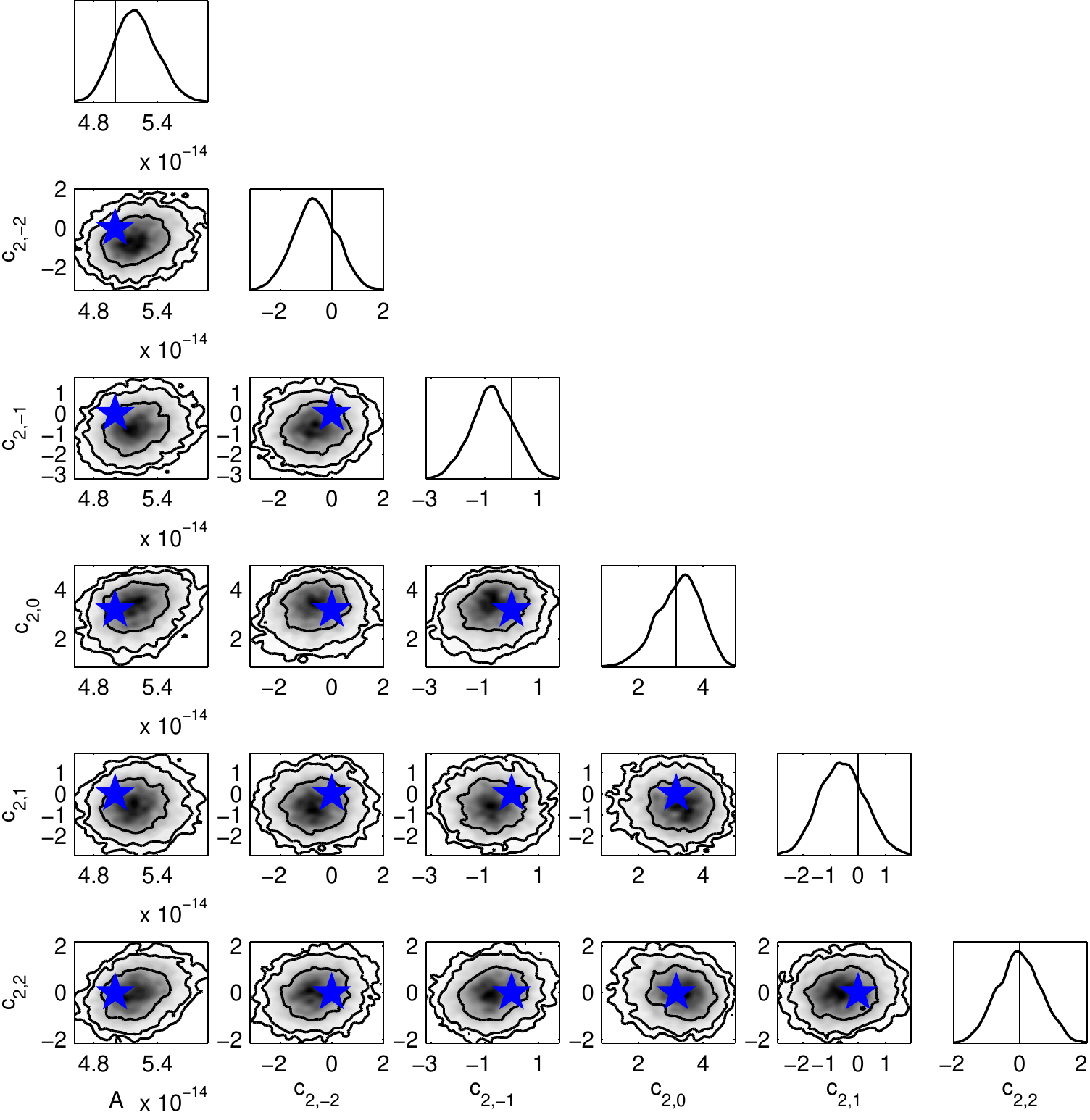}} 
   \subfloat[]{\incgraph{0}{0.5}{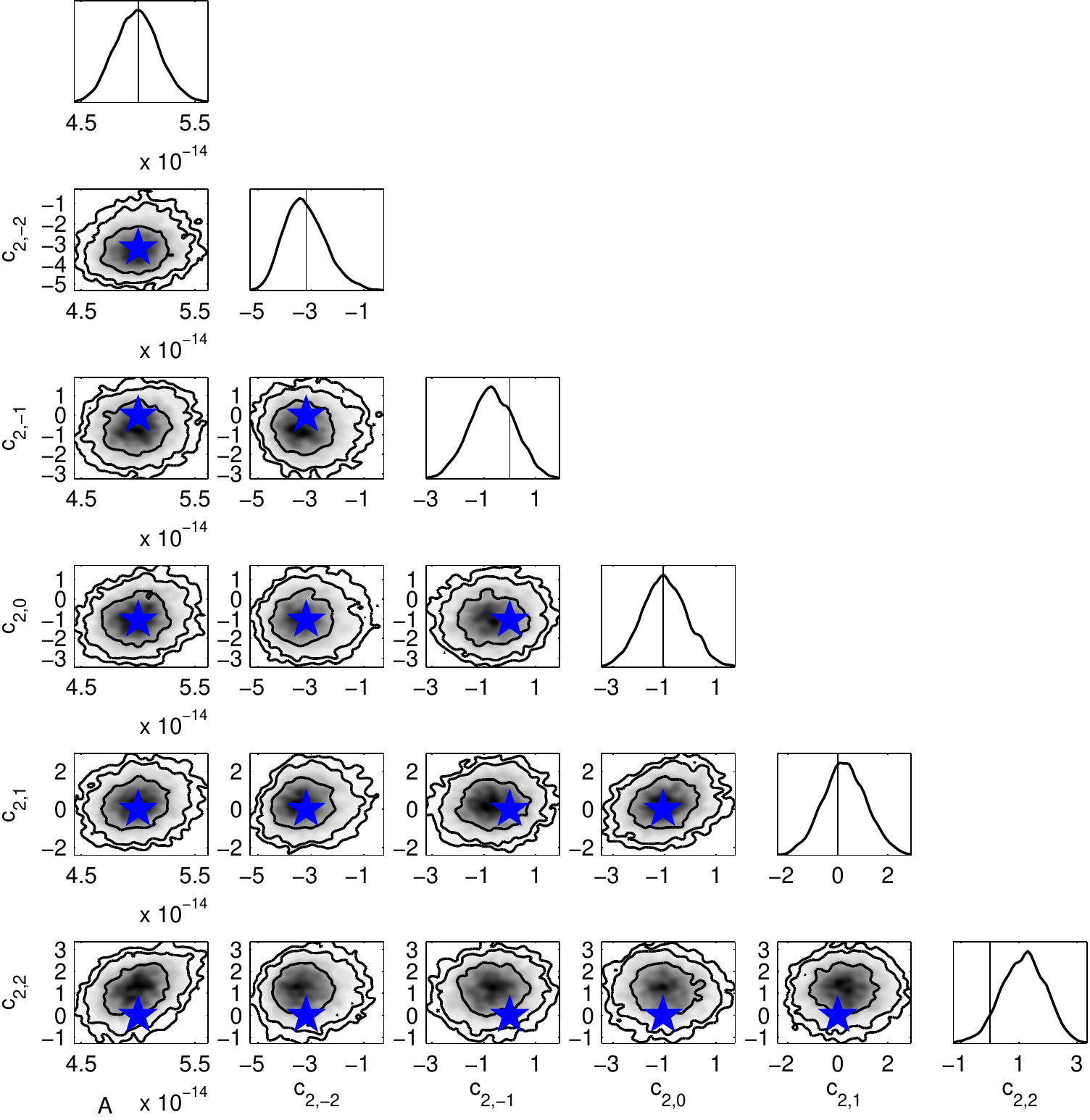}} \\
   \subfloat[]{\incgraph{0}{0.5}{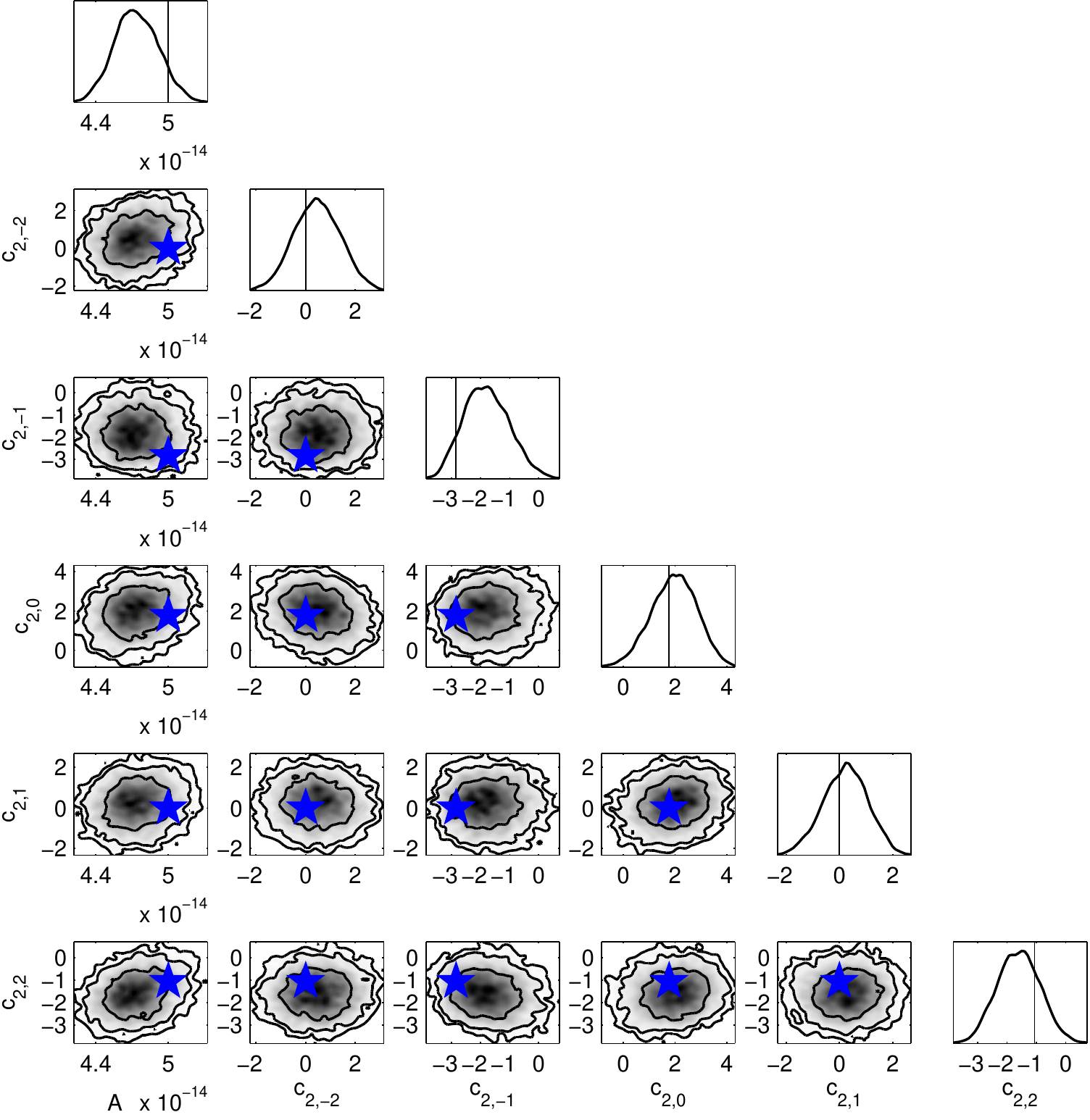}}
   \subfloat[]{\incgraph{0}{0.5}{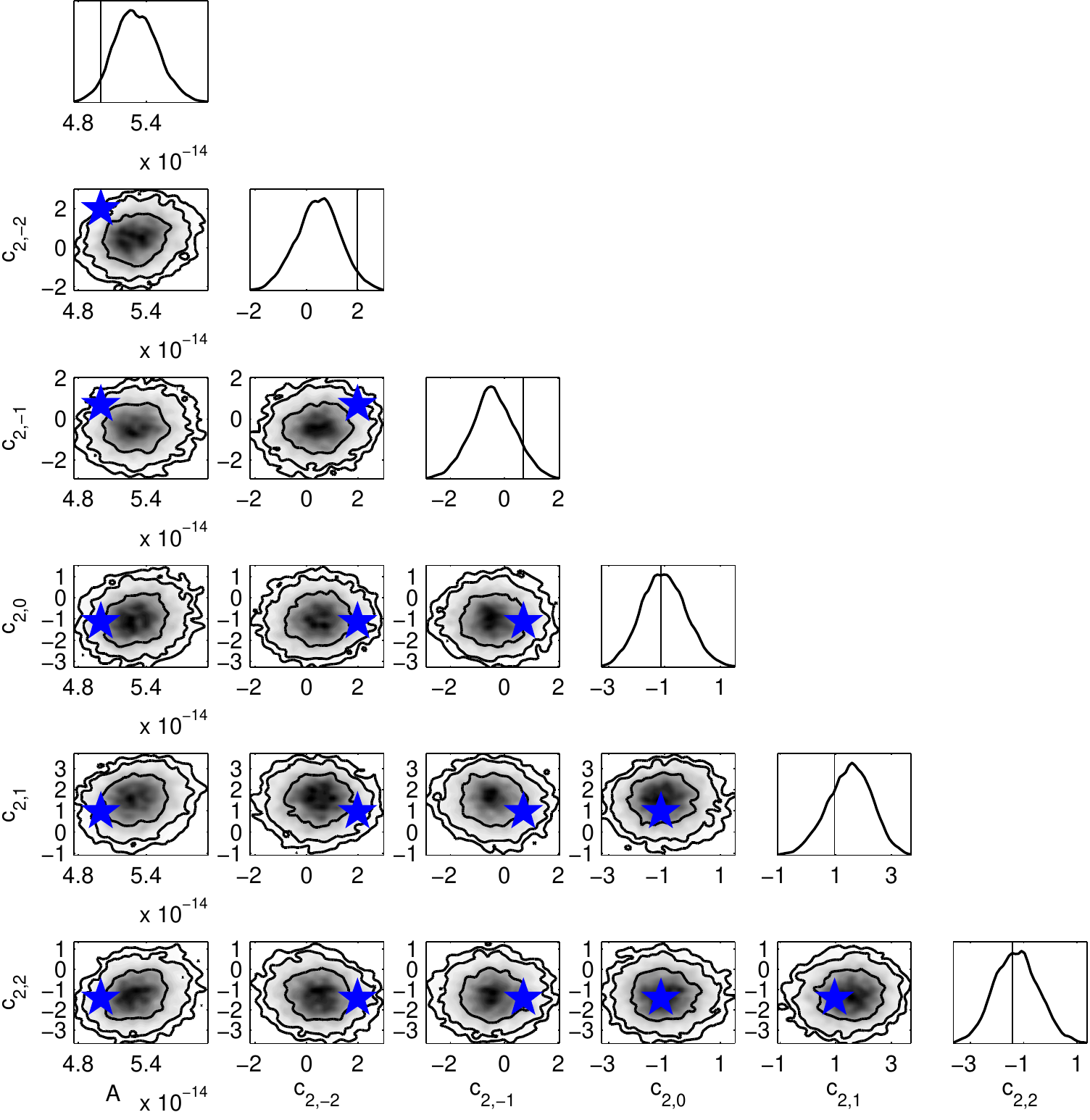}}
   \caption{\label{fig:Quad_Diff_Dist}The results of constraining the quadrupole-anisotropy coefficients, $c_{2,m}$ in datasets containing quadrupole anisotropies with a variety of configurations. The GW-source distributions corresponding to the injected anisotropies in each of these datasets are visualised in Figure \ref{fig:GWanisbkgrd_Quad_Dists}. The parameters of the injected anisotropies are (a) $\{c_{0,0}=2\sqrt\pi;\;c_{2,0}=4\sqrt{\pi/5}\}$ (b) $\{c_{0,0}=2\sqrt\pi;\;c_{2,-2}=-0.9 c_{0,0};\;c_{2,0}=-0.3 c_{0,0}\}$ (c) $\{c_{0,0}=2\sqrt\pi;\;c_{2,-1}=-0.8 c_{0,0};\;c_{2,0}=0.5 c_{0,0};\;c_{2,2}=-0.3 c_{0,0}\}$ (d) $\{c_{0,0}=2\sqrt\pi;\;c_{2,-2}=0.56 c_{0,0};\;c_{2,-1}=0.20 c_{0,0};\;c_{2,0}=-0.32 c_{0,0};\;c_{2,1}=0.27 c_{0,0};\;c_{2,2}=-0.40 c_{0,0}\}$.}
 \end{figure*}

As in the dipole case, we test the robustness of the generalised anisotropic formalism by generating datasets with quadrupole anisotropies in a variety of configurations. We begin with the power contained in one quadrupole moment, then allow all the $c_{lm}$'s to be non-zero. The GW-source distribution for each of the tested configurations is shown in Figure \ref{fig:GWanisbkgrd_Quad_Dists}. As can be seen in Fig.\ \ref{fig:Quad_Diff_Dist}, the generalised anisotropic-search pipeline recovers anisotropy coefficients which are consistent with the injected values. However the dimensionality is significantly larger than an isotropic search, and even in these high SNR datasets the posterior probability distributions are consistent with a wide range of possible anisotropies. This is partly due to the final three datasets having sub-maximal quadrupole anisotropy.

Finally, we repeat the analysis of the maximal-quadrupole dataset (with $P(\theta,\phi)=3\cos^2\theta$), but relax the assumption of quadrupole-anisotropy in the analysis. The results of this analysis are shown in Figure \ref{fig:Quad_Arb_Anis}, where we have accurately recovered the nature of the injected anisotropy. However, as observed in the case where we performed an arbitrary-anisotropy search on a pure-dipole dataset, the increased dimensionality of the arbitrary-anisotropy search widens the recovered posterior distributions to be consistent with a wide range of anisotropies.

\subsection{Arbitrary anisotropy}\label{sec:Arb_Anis_Section}

\subsubsection{Arbitrary anisotropy search, and a physical prior}

\begin{figure*}
   \subfloat[]{\incgraph{0}{0.32}{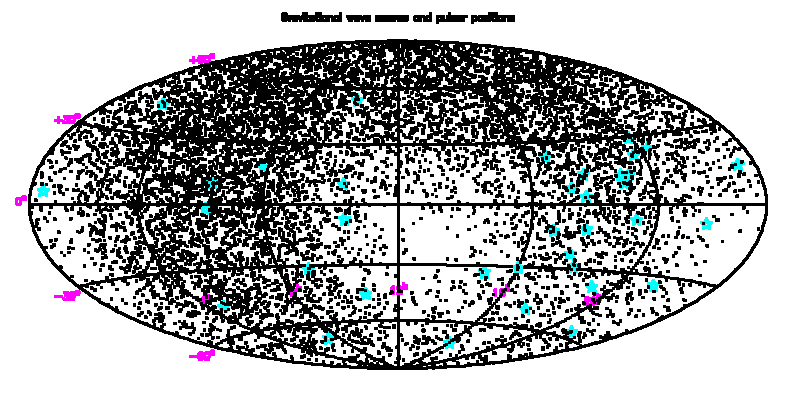}} 
   \subfloat[]{\incgraph{0}{0.32}{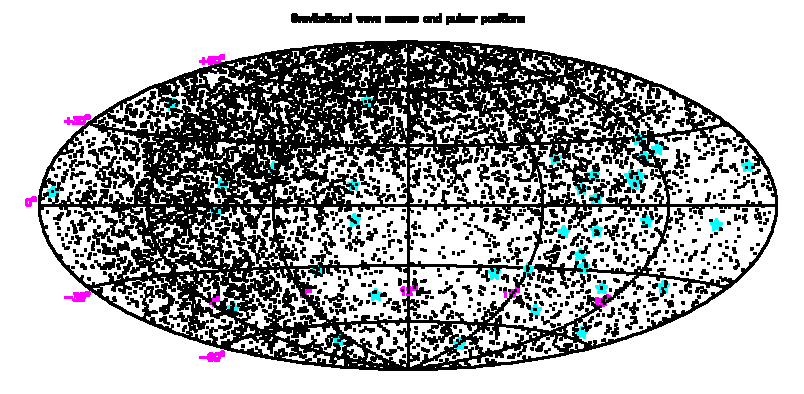}} 
   \subfloat[]{\incgraph{0}{0.32}{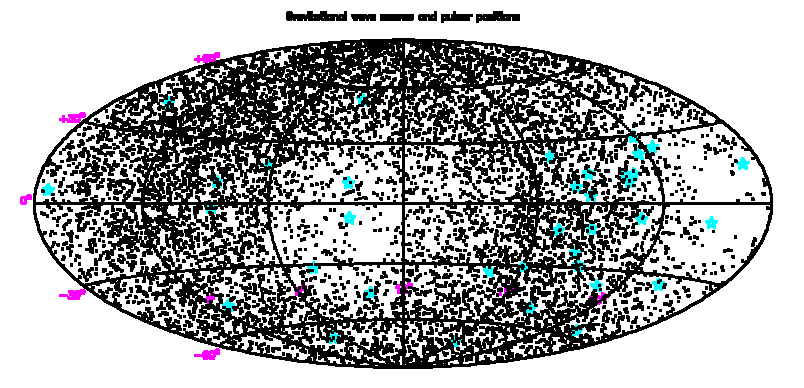}} 
   \caption{\label{fig:Arb_Dist_SkyMaps}In (a) we show the skymap for the distribution of GW sources implied by a GWB with anisotropy-coefficients $\{c_{0,0}=2\sqrt\pi;\; c_{1,-1}=-0.4 c_{0,0};\; c_{1,0}=0.4 c_{0,0};\; c_{2,0}=0.20 c_{0,0};\; c_{2,2}=-0.45 c_{0,0}\}$. In (b) we plot the skymap implied by the maximum a-posteriori anisotropic configuration from an analysis with our Bayesian serch-pipeline, where we have imposed a hard physical-prior (discussed in text). Finally, (c) shows the skymap implied by the maximum likelihood anisotropic configuration, where we see that, despite not having the extra constraining influence of the physical-prior, we are still able to adequately recover the distribution of GW sources.}
 \end{figure*}

We now test our search-pipeline on a dataset with an arbitrary anisotropy spread between monopole, dipole and quadrupole components. The anisotropy we have chosen has a positive probability distribution for the placement of GW-sources on the sky. The coefficients characterising the anisotropy are $\{c_{0,0}=2\sqrt\pi;\; c_{1,-1}=-0.4 c_{0,0};\; c_{1,0}=0.4 c_{0,0};\; c_{2,0}=0.20 c_{0,0};\; c_{2,2}=-0.45 c_{0,0}\}$, with all other anisotropy coefficients set to zero. The sky-map for the distribution of GW-sources on the sky corresponding to these anisotropy coefficients is shown in Fig.\ \ref{fig:Arb_Dist_SkyMaps}(a).

\begin{figure*}
   \incgraph{0}{1.0}{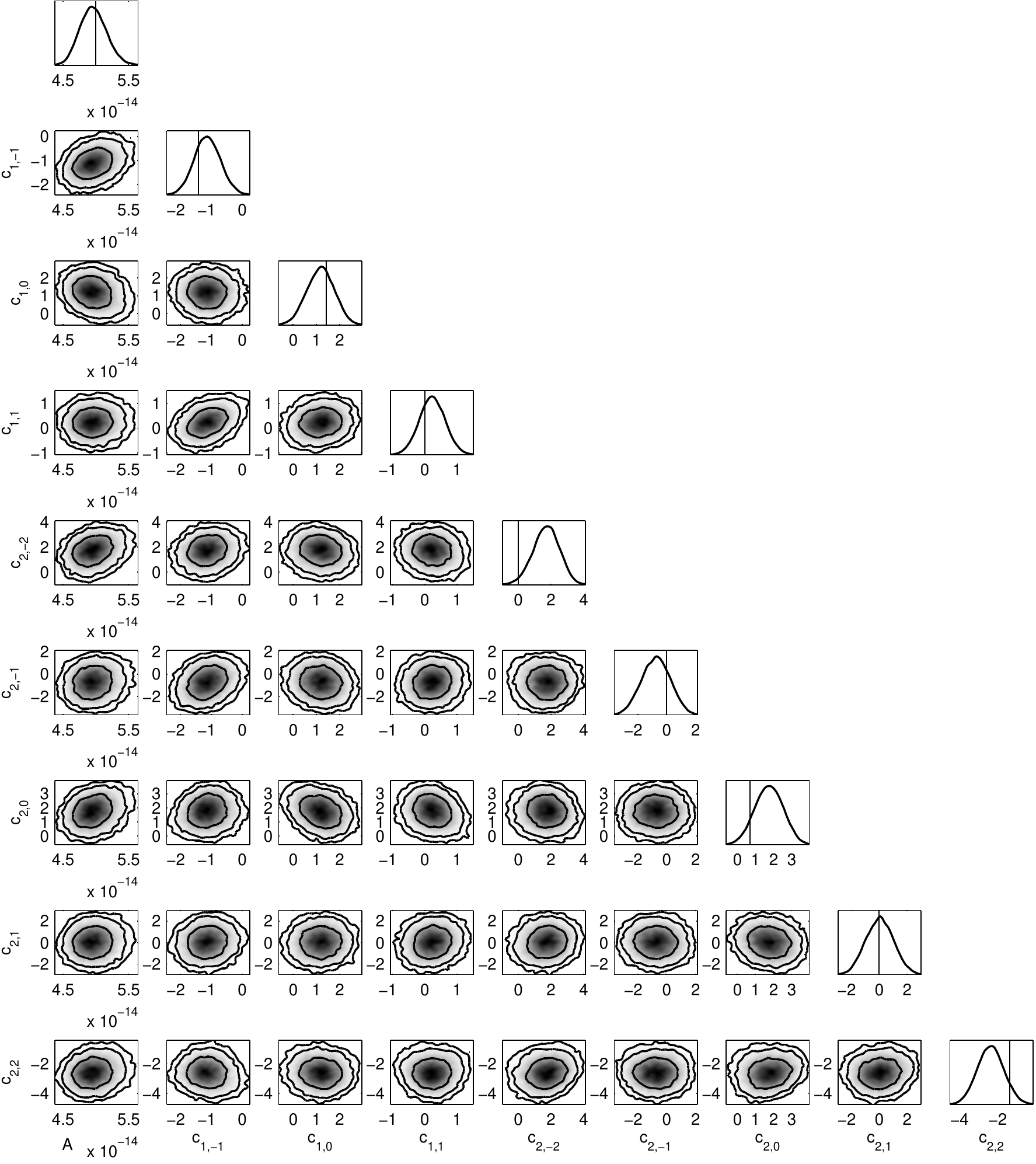}
   \caption{\label{fig:ArbAnis_WideFlatPrior}A full anisotropic-search on an arbitrary-anisotropy dataset. The details of the dataset are $\{c_{0,0}=2\sqrt\pi;\; c_{1,-1}=-0.4 c_{0,0};\; c_{1,0}=0.4 c_{0,0};\; c_{2,0}=0.20 c_{0,0};\; c_{2,2}=-0.45 c_{0,0}\}$. The prior is wide and flat, where $c_{l,m}\in[-5,5]$, such that all anisotropy-coefficient combinations are a-priori equally likely.}
 \end{figure*}

Taking a wide flat prior on the anisotropy coefficients as before, the results of a search are shown in Fig.\ \ref{fig:ArbAnis_WideFlatPrior}. The injected anisotropy coefficients all lie within the $2\sigma$ contours of the posterior distributions for the recovered coefficients. However, as commented on when the pure dipole and quadrupole datasets were subjected to an arbitrary-anisotropy search, the dimensionality degrades the precision with the coefficients can be recovered, so that the posterior is consistent with a wide range of anisotropic distributions.

\begin{figure*}
   \incgraph{0}{1.0}{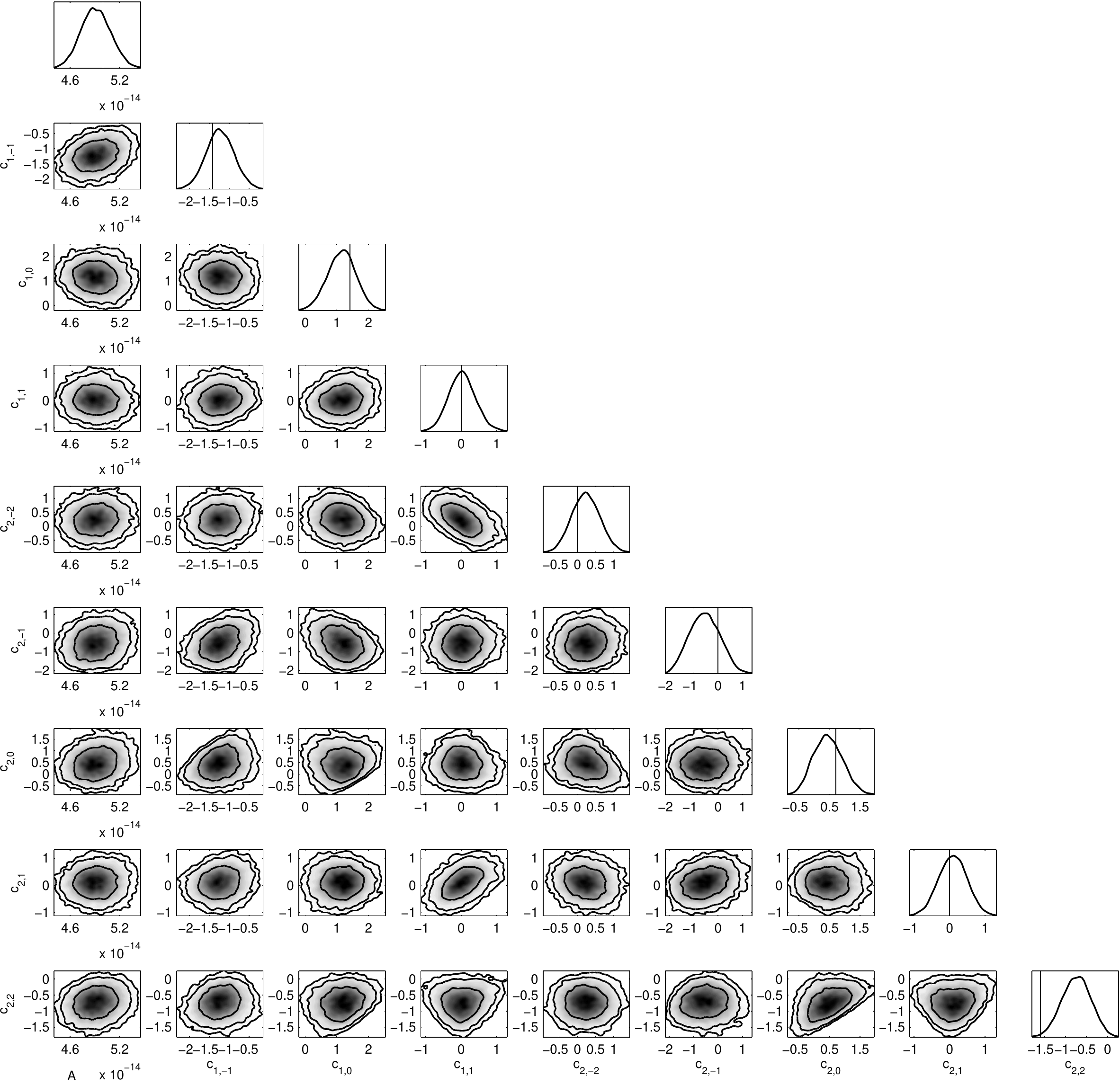}
   \caption{\label{fig:ArbAnis_WidePhysPrior}A full anisotropic search on the same arbitrary-anisotropy dataset as in Figure \ref{fig:ArbAnis_WideFlatPrior}. The anisotropy coefficients used to generate the dataset are $\{c_{0,0}=2\sqrt\pi;\; c_{1,-1}=-0.4 c_{0,0};\; c_{1,0}=0.4 c_{0,0};\; c_{2,0}=0.20 c_{0,0};\; c_{2,2}=-0.45 c_{0,0}\}$. The prior is the only physically-meaningful prior on possible anisotropic distributions. This corresponds to only sampling in regions where $\sum_{l,m}c_{lm}Y_{lm}(\hat\Omega)\geq 0,\;\forall\;\hat\Omega$.}
 \end{figure*}

We now address this with a more physically-motivated prior choice. Until now we have assumed that all combinations of anisotropy-coefficients are a-priori equally likely. However, when generating these anisotropic-datasets we have had to take into account that some combinations do not represent physical anisotropies, since they fail the condition, Eq.~(\ref{eq:probconstraint}), that the PDF for the distribution of GW-sources on the sky is positive at all sky-locations,
\begin{equation}
P(\hat\Omega)\propto \frac{dN}{d\hat\Omega} \propto \sum_{l,m}c_{lm}Y_{lm}(\hat\Omega)\geq 0,\quad\forall\;\hat\Omega.
\end{equation}
Therefore, if we are to correctly perform parameter estimation on anisotropic datasets, this condition is the only physically meaningful prior on the anisotropy-coefficients. We impose it by setting the likelihood of any combination of $c_{l,m}$ which fails this condition to be very low, so as to prevent sampling in these regions. In practice this is achieved by pixelating the sky with a $20\times 20$ grid in $[\cos\theta,\phi]$, and testing the condition in each pixel. If the condition is not met in every pixel, then the sample is rejected. The result of such an application of this prior on the previously analysed arbitrary-anisotropy dataset is shown in Fig.\ \ref{fig:ArbAnis_WidePhysPrior}. For the most part, the recovered posterior distributions are consistent with the injected anisotropy. The exception is for the $c_{2,2}$ coefficient, where the injected amplitude lies in the tails of the recovered posterior distribution. The reason for this apparent inconsistency is actually quite straightforward. If we attempt to generate a dataset with all anisotropy coefficients as before, except for $c_{2,2}$, which we lower from $-0.45c_{0,0}$ to $-0.47c_{0,0}$, then the GWanisobkgrd plugin fails, since this PDF for the placement of GW-sources on the sky is negative in some sky-locations. Thus the reason that the injected $c_{2,2}$ amplitude is in the tails of the recovered posterior distribution is because this value is a-priori disfavoured, and very close to the boundary of non-physicality.

Despite the apparent inconsistency of the $c_{2,2}$ coefficient, we are still able to accurately recover the distribution of GW-sources on the sky. In Figure \ref{fig:Arb_Dist_SkyMaps} we compare the sky-maps of GW-background realisations for the injected anisotropy, maximum-a-posteriori anisotropy of an analysis with the physical-prior imposed on the $c_{l,m}$ coefficients, and the maximum-likelihood anisotropy where the relative power in $l=0$ has been boosted to ensure physicality over the entire sky. As can be seen, the anisotropic-search pipeline has correctly recovered the features of this background. Even the maximum-likelihood anisotropic configuration, despite being unphysical, can have the relative power of its isotropic component boosted to ensure physicality over the entire sky, producing a GW source-distribution which is adequately close to the injected distribution.

\subsubsection{Anisotropy misfits}

\begin{table*}
\caption{\label{tab:Anis_Misfits}Testing the ability of the anisotropic-search pipeline to identify the nature of the injected anisotropy. We subject datasets with different anisotropies (isotropic, dipole, quadrupole, arbitrary) to isotropic, dipole, quadrupole and arbitrary searches. We are interested in evidence values, and so we adopt wide flat priors on the anisotropy coefficients ($c_{l,m} \in [-5,5]$) since the physical-prior discussed in the previous sub-section is actually very informative, truncating regions of high-likelihood. The details of the injected arbitrary-anisotropy dataset are $\{c_{0,0}=2\sqrt\pi;\; c_{1,-1}=-0.4 c_{0,0};\; c_{1,0}=0.4 c_{0,0};\; c_{2,0}=0.20 c_{0,0};\; c_{2,2}=-0.45 c_{0,0}\}$. We quote log-Bayes' factors for the anisotropic models versus the isotropic model for each combination of injected anisotropy and model used to test for anisotropy.}
\begin{ruledtabular}
\begin{tabular}{cccccc}
& & \multicolumn{4}{ c }{Injected anisotropy, \{details\}} \\ 
& & $l=0$ & $l=0,1$ & $l=0,2$ & $l=0,1,2$ \\ 
& & $\{c_{0,0}=2\sqrt\pi\}$ & $\{c_{0,0}=2\sqrt\pi,c_{1,0}=2\sqrt{\pi/3}\}$ & $\{c_{0,0}=2\sqrt\pi,c_{2,0}=4\sqrt{\pi/5}\}$ & \{details in caption\} \\ \cline{1-6}
\multicolumn{1}{ c }{\multirow{1}{*}{Tested} } &
\multicolumn{1}{ c }{$l=0$, $\{A\}$} & 0.0 & 0.0 & 0.0 & 0.0 \\ 
\multicolumn{1}{ c  }{\multirow{1}{*}{anisotropy,}} &                       
\multicolumn{1}{ c }{$l=0,1$, $\{A,c_{1,m}\}$} & -5.8 & 2.2 & -2.7 & 5.8 \\ 
\multicolumn{1}{ c  }{\multirow{1}{*}{\{search} } &
\multicolumn{1}{ c }{$l=0,2$, $\{A,c_{2,m}\}$} & -5.6 & -4.4 & 2.8 & 8.9 \\
\multicolumn{1}{ c  }{\multirow{1}{*}{parameters\}}} &                      
\multicolumn{1}{ c }{$l=0,1,2$, $\{A,c_{l,m}\}$} & -10.4 & -3.9 & -2.3 & 9.7 \\
& \multicolumn{1}{ c }{True, $\{A\}$} & 0.0 & 8.2 & 7.7 & 17.6  \\
\end{tabular}
\end{ruledtabular}
\end{table*}

We now test the ability of the anisotropic-search pipeline to recover the \textit{nature} of the GW-background anisotropy. Taking isolated SNR $=10$ realisations of isotropic, dipole, quadrupole and arbitrary anisotropy datasets, we computed the evidence for isotropic, dipole, quadrupole and arbitrary models. We also computed the evidence for the true, injected anisotropy, searching only over the amplitude of the background. The results are shown in Table \ref{tab:Anis_Misfits}, where all numbers indicate log-Bayes' factors relative to an isotropic model. For these evidence calculations, we adopt the flat priors on the anisotropy coefficients. Although in practice we should be imposing the physicality conditions, we have found that the physical-prior studied in the previous sub-section is actually highly informative, and can truncate regions of high-likelihood. The anisotropic configurations we have studied are close to maximal, which means the maximum-likelihood parameter regions will be close to non-physicality. Taking the flat priors can therefore help improve the anisotropic evidence, since the high likelihood region stretches beyond the boundary imposed by physicality. Thus, for these cases, the unphysical evidence seems to be a better statistic to use to distinguish models than the physical evidence. Further investigations will be required to determine if this is true more generally, in particular for cases in which the anisotropy is weak. Interpreting the unphysical distributions on the coefficients is more difficult, since they give rise to negative probabilities for certain regions of the sky. However, increasing the relative power of the isotropic $l=0$ mode can always be used to make the distribution physical, and provide source distributions which are adequately close to the injected distribution. The effect of the unphysical anisotropic coefficients, $c_{lm}$, on our search is to allow the GWB-correlations deviate by more from the Hellings and Downs curve, allowing some of the noise to be absorbed into these coefficients. This seems to help for the limited number of cases we have considered but further work is required to understand which prior works better in more general situations.

\begin{figure}
   \incgraph{0}{0.5}{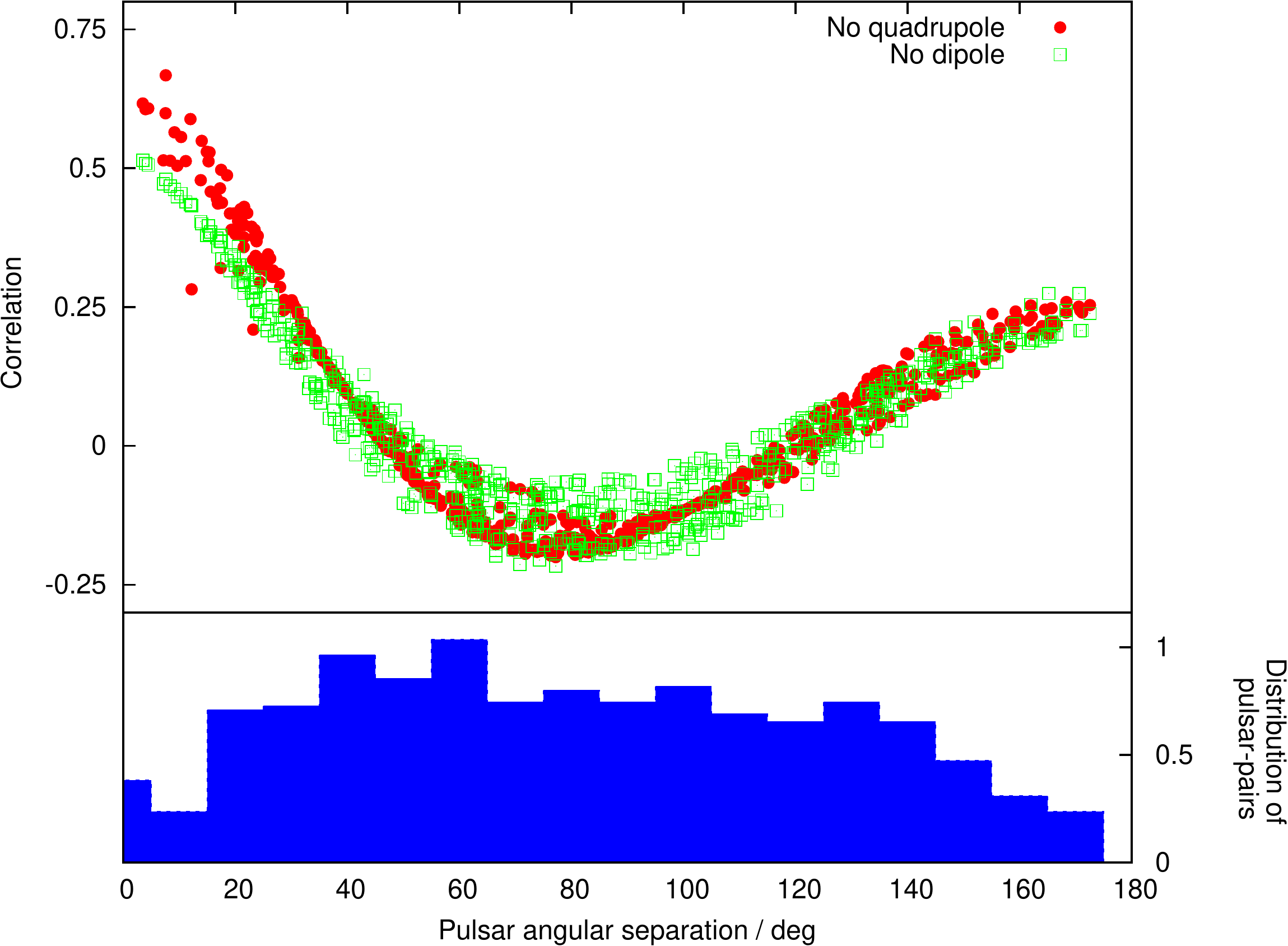}
   \caption{\label{fig:Arb_QuadDominates}We show the angular correlations for pulsar-pairings with non-zero angular-separation in an arbitrary-anisotropy dataset, for the cases where we ignore the contribution from the dipole anisotropy and when we ignore the quadrupole anisotropy. In the region of pulsar angular-separation where there are many distinct pulsar-pairings, the quadrupole-only anisotropy dominates the deviation of correlations from the isotropic Hellings and Downs curve. This may explain why a quadrupole-only model is almost as good as an arbitrary-anisotropy model when describing the arbitrary-anisotropy dataset. Note that zero angular-separation pairings are not shown on the angular-correlation sub-plot, but are included in the pulsar-pairing histogram.}
 \end{figure}

Referring to Table \ref{tab:Anis_Misfits}, the isotropic model is highly favoured when describing the isotropic dataset. The true dipole anisotropy is substantially favoured over an isotropic model when describing the dipole-dataset, and likewise the quadrupole anisotropy is substantially favoured over an isotropic model when describing the quadrupole-dataset. Finally, in the model-comparison of the arbitrary anisotropy dataset, any form of anisotropy is favoured over the isotropic model. However, an arbitrary anisotropic-search is favoured over both the dipole and quadrupole searches, but in the latter case the degree to which the arbitrary anisotropic-search is favoured is inconclusive. We suspect the reason for this can be understood in terms of how the different anisotropies affect the scatter of correlations around the isotropic Hellings and Downs curve, particularly for pulsar angular-separations where there are many pulsar-pairings. This is shown in Fig.\ \ref{fig:Arb_QuadDominates}, where we see that in the region where there are many distinct pulsar-pairings the quadrupole-anisotropy induces a larger scatter and deviation from the isotropic Hellings and Downs curve than the dipole-anisotropy. Hence, in this case, a quadrupole-only model is almost as good as an arbitrary-anisotropy model when describing the arbitrary-anisotropy dataset.


\section{Conclusions}\label{sec:conclusions}

We have systematically investigated the robustness and veracity of a Bayesian search-pipeline capable of probing anisotropic gravitational-wave backgrounds (GWBs) using pulsar-timing arrays (PTAs). If we have some form of continuous anisotropy in the energy density of such a GWB, or anisotropy through sparseness of sources comprising our background, then the assumption of isotropy may no longer be an appropriate model with which to perform searches. Recent work by \citet{cornish-sesana-2013}, and \citet{chiara-paper} has focused on developing new formalisms and techniques with which to generalise the current search methodologies, which all fall within the categories of either stochastic isotropic searches or single-source searches.

We employed the formalism of \citet{chiara-paper}, which generalises the cross-correlation of GWB-induced TOA-deviations between pulsars to anisotropic backgrounds. In an isotropic GWB this cross-correlation has a unique, distinctive signature, being a function only of the angular separation between pulsars on the sky. It is commonly referred to as the Hellings and Downs curve \citep{hellings-downs-1983}. In the more generalised formalism, the distribution of GWB power is decomposed as a function of spherical-harmonics, and a set of correlation basis-functions uniquely defined for a particular PTA configuration, although the aforementioned authors have identified useful analytic expressions for these basis-functions up to quadrupole order. Unsurprisingly, the monopole solution, describing the induced correlation for the isotropic mode, is the usual Hellings and Downs curve. Crucially, the correlation basis-functions for dipole anisotropies and above are no longer independent of the absolute sky-location of the pulsar, being strongly dependent on the placement of the pulsars relative to the distribution of GWB-power.

To thoroughly investigate this formalism, we have developed a suite of new plugins for the popular pulsar-timing analysis software package, \textsc{Tempo2}. This permits the user to have complete control over the generation of realistic-format datasets, including specifying the array of pulsars, the various noise processes affecting these pulsars, and the ability to inject a GWB signal into these simulated datasets. We have generalised this to inject continuous anisotropies into the GWB signals, and also to permit the inclusion of non-Einsteinian polarisation modes in the GW-signals.

By generalising the cross-correlation of pulsar-timing residuals in a time-domain Bayesian search, we have probed the level of anisotropy in the distribution of GW source-populations making up many background realisations. In particular, we have found that our pipeline can infer the direction of a strong dipole-anisotropy in a loud GWB. Not only that, but by performing Bayesian evidence evaluation using the \textsc{MultiNest} package, such a strong dipole-anisotropy can be favoured over the standard isotropic search, where we see that the latter is no longer an appropriate model to use. By evaluating the evidence of datasets with the anisotropy fixed at the injected levels, and subsequently comparing to the isotropic evidence, we find that the best-case scenario for having decisive evidence for a strong-dipole anisotropy occurs when the induced rms-residual of the GWB is approximately five times greater than the noise rms-residual. When a more general search over the dipole anisotropy-coefficients is performed, the log-Bayes factors are degraded, but substantial evidence for the anisotropic signal may become visible at SNRs $\geq 10$. We performed a similar analysis with uncorrelated red-noise affecting the pulsars, finding a similar level of required GWB-loudness in order to begin having convincing evidence of anisotropy.

We extended our search to datasets with injected GWBs containing quadrupole anisotropies in a variety of different configurations. Again, we found that the search-pipeline was able to recover posterior probability distributions for the anisotropic-decomposition coefficients, finding agreement with injected values. We repeated our analysis of the best-case scenario for when we could have convincing evidence of anisotropy in the GWB, this time for the quadrupole-anisotropy. As before, in this absolute best-case scenario, where we fix the tested anisotropy to the injected values, we still require the rms-residual induced by the GWB to approximately five times greater than the noise rms-residual. When a more general search over the quadrupole anisotropy-coefficients is performed, the log-Bayes factors are again degraded, but substantial evidence for the anisotropic signal is possible at SNR $=10$, such that clear evidence would require SNRs $\geq 10$. Further work is needed here, in order systematically evaluate the Bayesian evidence at different GWB loudnesses without suffering from the curse of high-dimensionality, which can have a detrimental effect when comparing to isotropic models.

Finally, we tested our search-pipeline on a dataset with the distribution of GWB-power spread over monopole, dipole and quadrupole. While the recovered maximum likelihood anisotropy produces a distribution of GW sources which adequately mimics the injected distribution, the recovered probability distribution of anisotropy-coefficients is rather wide since we are now searching over nine dimensions. Thus, we could conceivably be consistent with a wide range of anisotropic configurations. Hence we propose a hard prior on the anisotropy-coefficients, corresponding to the physicality of the implied probability distribution for the placement of sources on the sky. If this PDF happens to be negative anywhere on the sky for a proposed set of anisotropy-coefficients, then this sample is discarded in the MCMC process. This is actually quite an informative prior, such that if we are studying a dataset with an anisotropy close to non-physicality, then the prior may actually truncate regions of high likelihood in parameter space. This could conceivably have a negative impact on evidence evaluation, however it is a very useful tool in parameter-estimation. The maximum a-posteriori anisotropy when this physicality-prior is imposed shows a greater match to the injected anisotropy. We intend to study the effects of this prior in future work.

The search for angular structure in the GWB will be the obvious next step in GWB PTA-searches, permitting us to investigate the clustering of sources or GW-hotspots. This paper has robustly tested an anisotropic stochastic search-pipeline on datasets with a continuous anisotropy in the angular-distribution of sources constituting the injected background. We intend to follow up this first study with a further investigation of the consequences of imposing a hard physical-prior, as well as applying this search-pipeline to datasets with GWBs consisting of realistic populations of inspiraling SMBHBs. This latter investigation will allow us to understand the effects of source-sparseness and background-finiteness (which occur at higher GW frequencies as the stochasticity of the strain signal breaks down) on the ability to constrain the properties of a GWB. The next several years should see rapid advances in the field of PTA searches for GWs, as we move beyond the standard paradigm of an isotropic search, develop faster, more robust algorithms, and perform the first combined search using the efforts and resources from all three PTAs.

\begin{appendix}
\section*{}

\begin{figure*}
   \subfloat[]{\incgraph{0}{0.5}{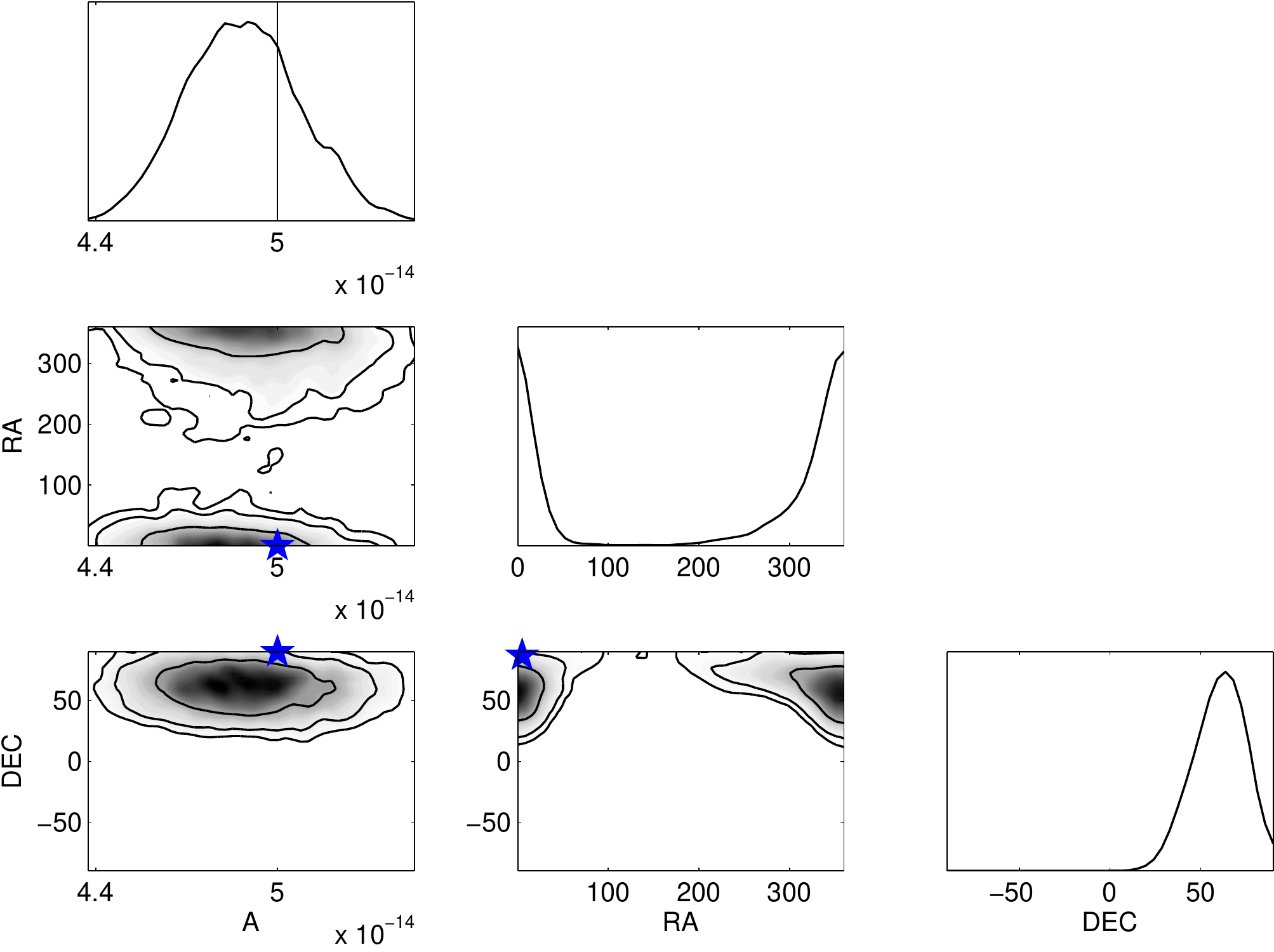}} 
   \subfloat[]{\incgraph{0}{0.5}{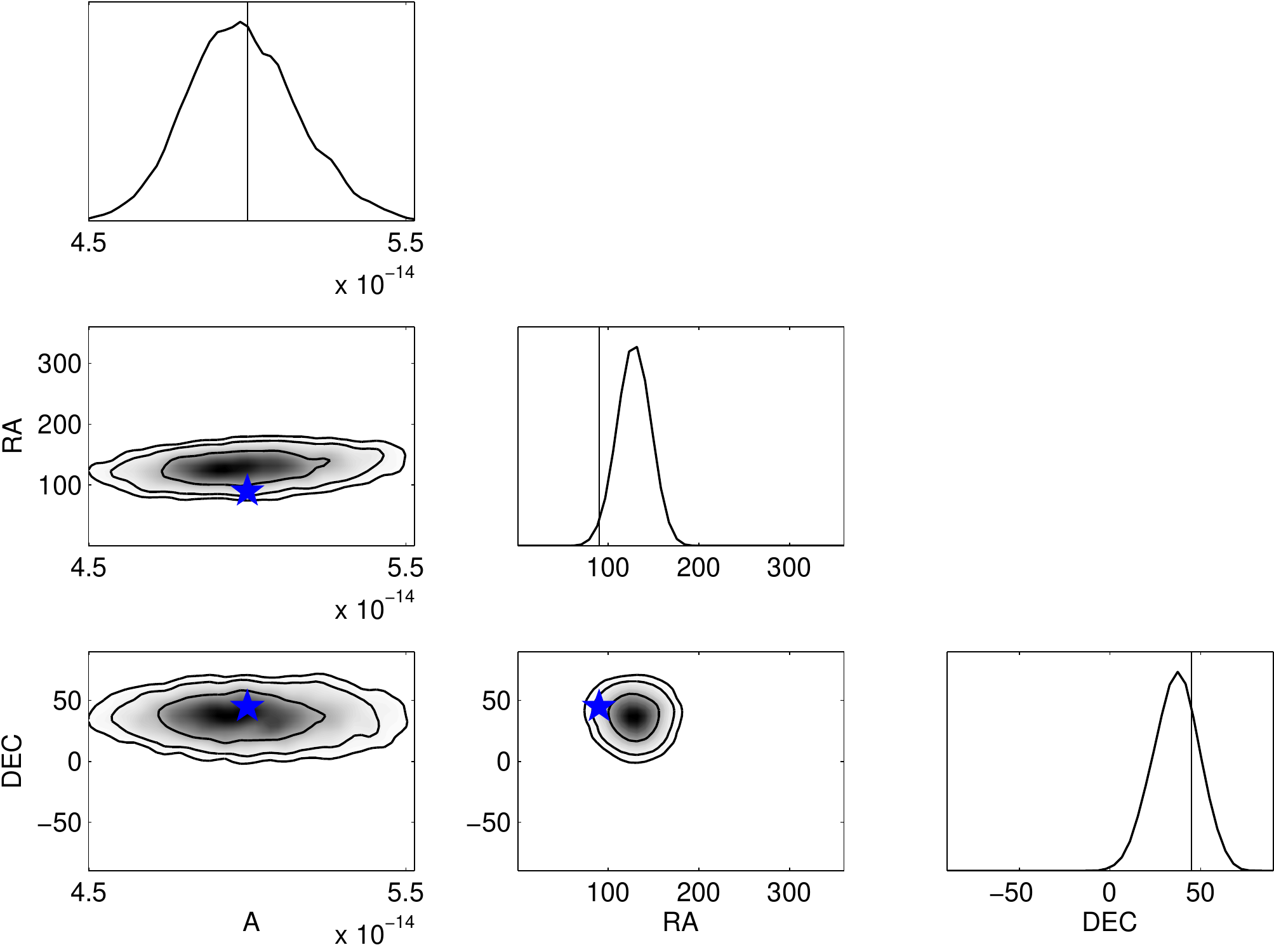}} \\
   \subfloat[]{\incgraph{0}{0.5}{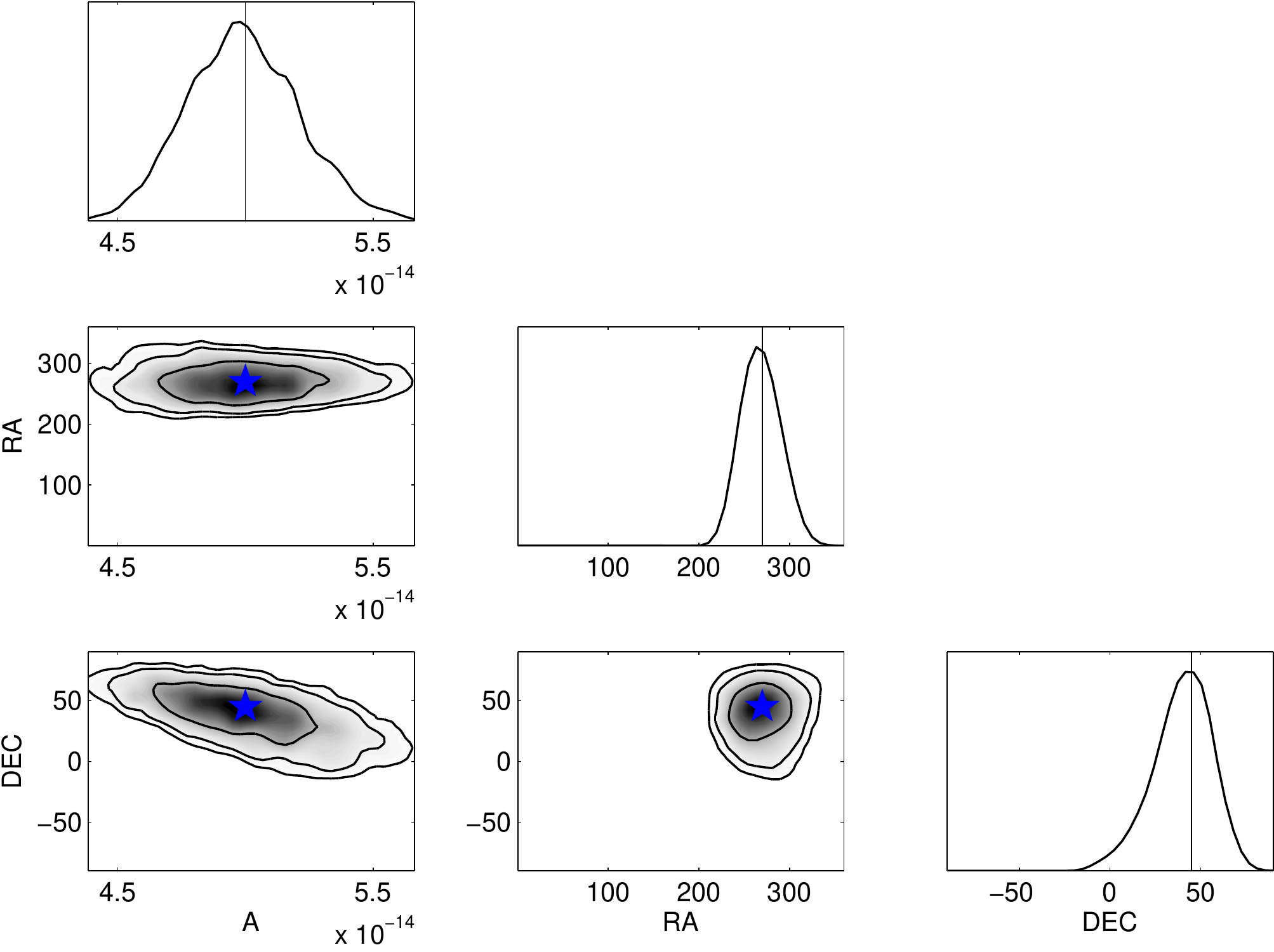}}
   \subfloat[]{\incgraph{0}{0.5}{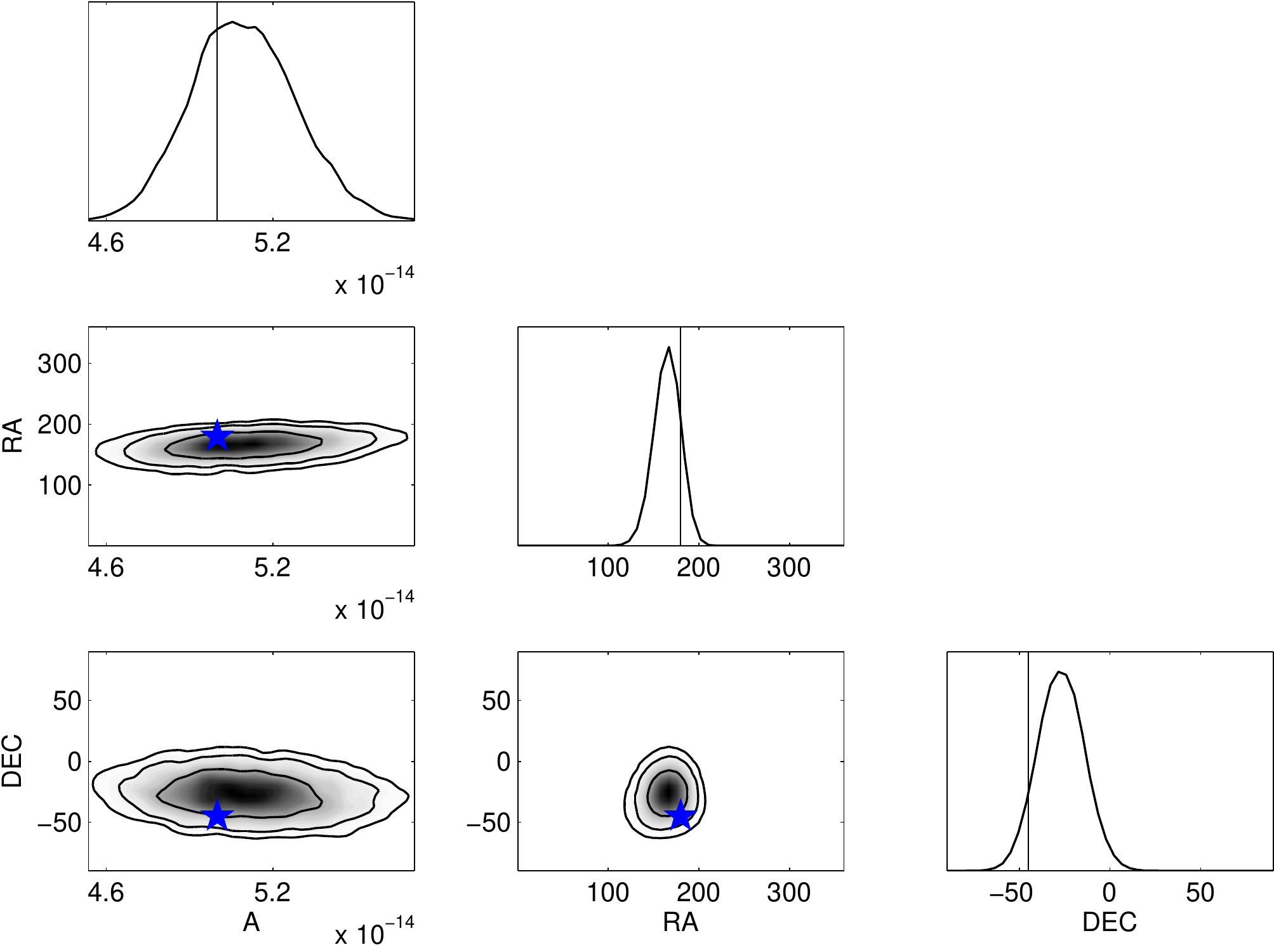}}
   \caption{\label{fig:DipDirectionRecon_3D}We repeat the analysis of the dipole-direction recovery, as in Fig.\ \ref{fig:DipDirectionRecon_4D}, but this time fix $\gamma$ to $13/3$. As expected, we are able to recover $A$ and the (RA, DEC) of the injected dipole without bias.}
 \end{figure*}

\begin{figure}
   \incgraph{0}{0.5}{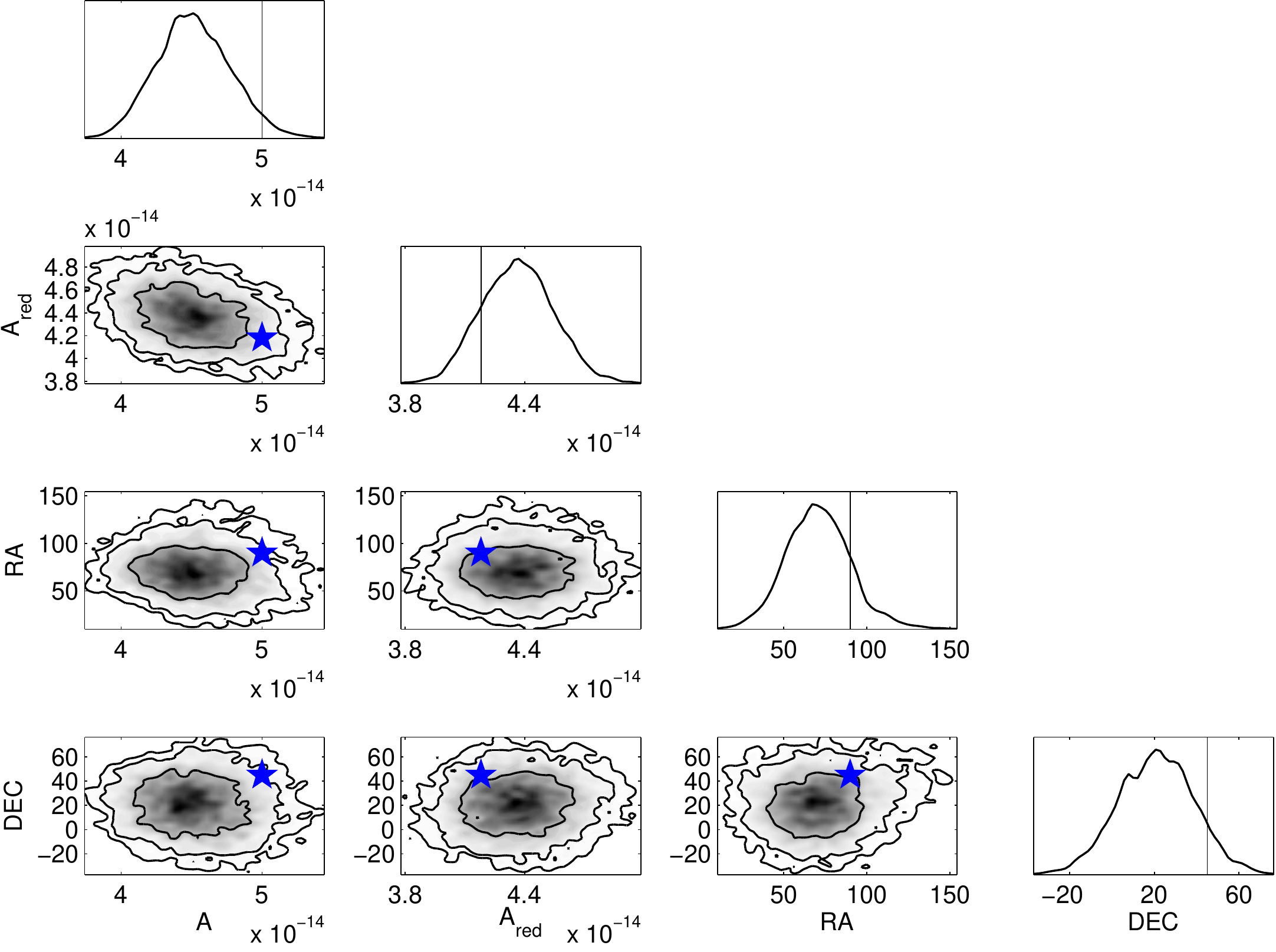}
   \caption{\label{fig:Dipole_RED_SNR10_GammaFixed}Reconstructing the amplitudes of the GWB and the uncorrelated red-noise spectrum when their spectral-slopes are fixed at the injected values of $13/3$ and $1.7$ respectively. The dataset is the same as in Fig.\ \ref{fig:Dipole_RED_SNR10}. By fixing the spectral-slope values, we can accelerate the analysis without biasing the dipole-direction reconstruction. Furthermore, the value of $\ln(\mathcal{B})$ in this $4$D case is only $\sim 0.8$ larger than in the full $6$D scenario.}
 \end{figure}

\begin{figure*}
   \incgraph{0}{1.0}{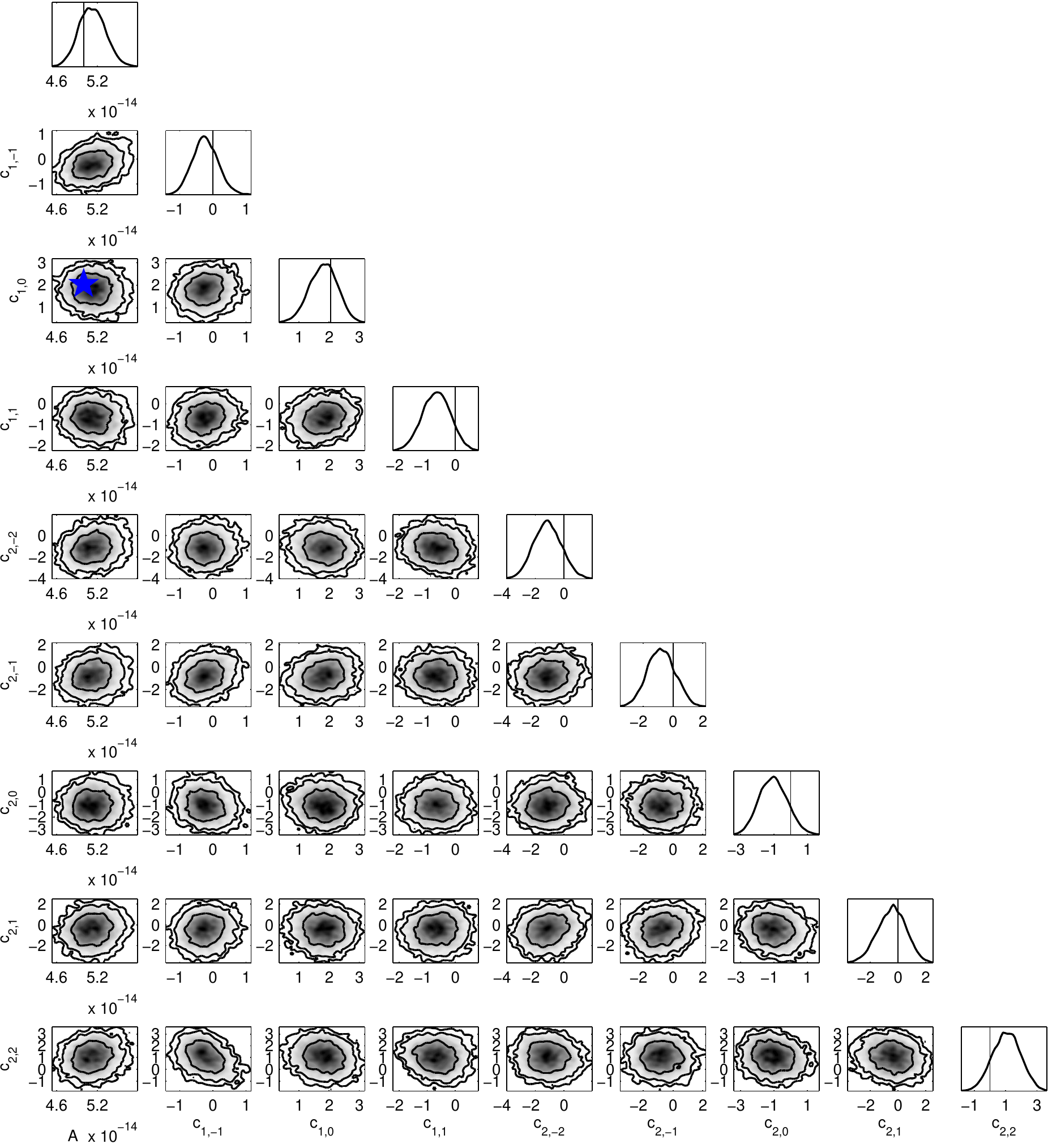}
   \caption{\label{fig:Dip_Arb_Anis}A search over all anisotropy coefficients, $c_{lm}$, up to, and including, quadrupole order. The injected anisotropy was maximal-dipole with $c_{0,0}=2\sqrt{\pi}$ and $c_{1,0}=2\sqrt{\pi/3}$, corresponding to a dipole-direction of RA$=0.0^{\circ}$, DEC$=90^{\circ}$. As can be seen, the anisotropic-search pipeline fully recovered all details of the injected anisotropy, as well as the amplitude of the GWB characteristic strain-spectrum, $A$.}
 \end{figure*}

\begin{figure*}
   \incgraph{0}{1.0}{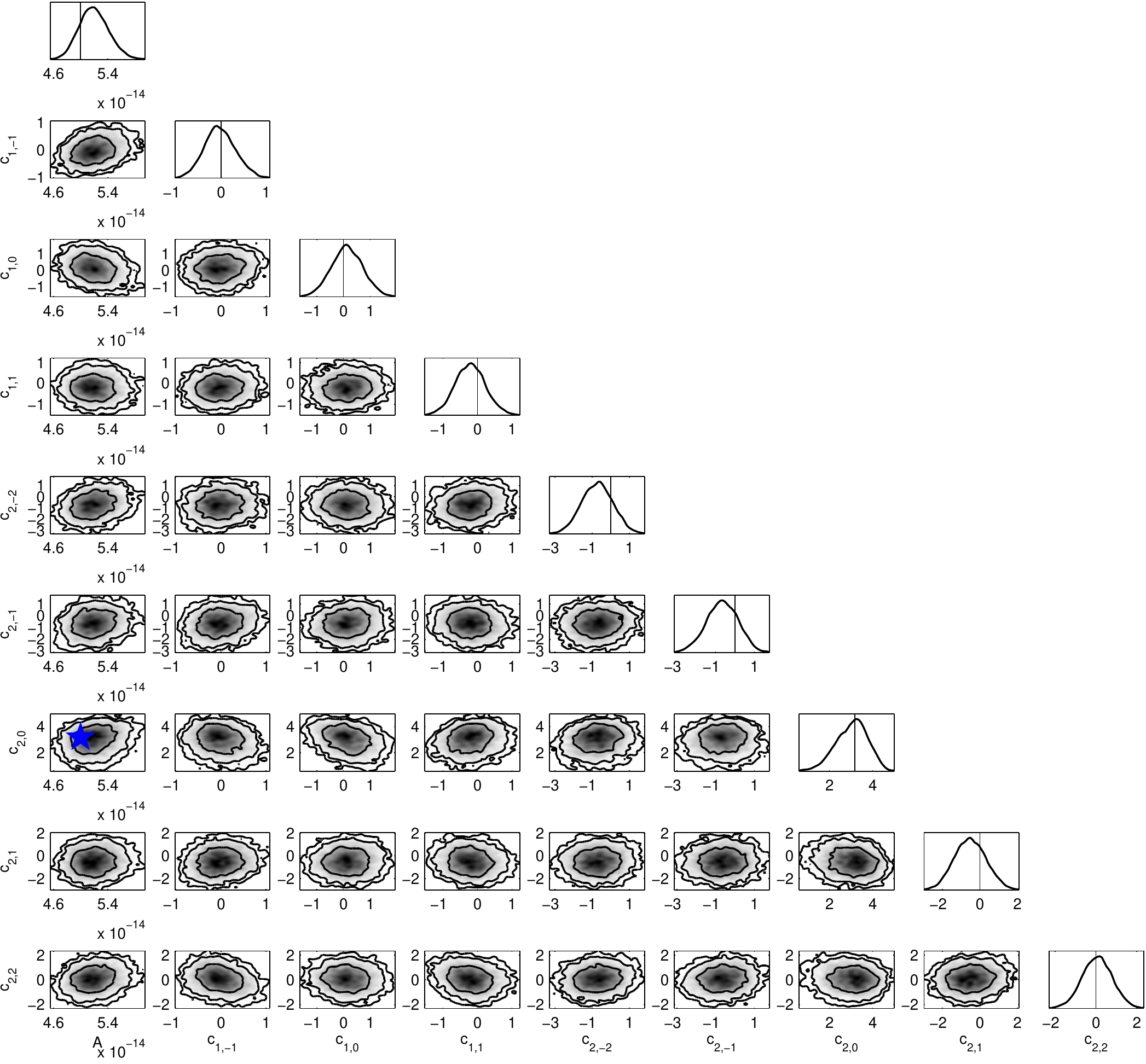}
   \caption{\label{fig:Quad_Arb_Anis}A search over all anisotropy coefficients, $c_{lm}$, up to, and including, quadrupole order. The injected anisotropy was maximal-quadrupole with $c_{0,0}=2\sqrt{\pi}$ and $c_{2,0}=4\sqrt{\pi/5}$, and all other dipole/quadrupole coefficients equal to zero. As can be seen, the anisotropic-search pipeline fully recovered all details of the injected anisotropy, as well as the amplitude of the GWB characteristic strain-spectrum, $A$.}
 \end{figure*}

\end{appendix}

\begin{acknowledgments}
We thank Chiara Mingharelli and Alberto Vecchio for many useful discussions and Alberto Sesana for supplying the data used to generate the figures in Section~\ref{sec:ExpectedAnisSection}. S.R.T is supported by the STFC, and acknowledges the support of the Royal Astronomical Society. J.R.G is supported by the Royal Society. This work was performed using the Darwin Supercomputer of the University of Cambridge High Performance Computing Service (http://www.hpc.cam.ac.uk/), provided by Dell Inc. using Strategic Research Infrastructure Funding from the Higher Education Funding Council for England. 
\end{acknowledgments}

\bibliography{pta_anis_refs}

\end{document}